\documentclass {article}
\usepackage {isolatin1,latexsym}
\begin {document}


\pagestyle {myheadings}
\markright{\large A new discrete view to quantum mechanics}

\newcommand{\FN}[1]{\footnote {#1} } 
\newcommand{\MBOX}[1]{\quad \mbox {#1} \quad}
\newcommand{\CC}[1]{#1^\ast} 
\newcommand{\HC}[1]{#1^\dagger} 
\newcommand{\BAR}[1]{\overline {#1}} 
\newcommand{\HCB}[1]{\widetilde{#1}}
\newcommand{\MV}[1]{{\bf #1}} 
\newcommand{\TR}[1] {\mathcal T( #1 )} 
\newcommand{\VAR} {\delta \! } 
\newcommand{\LAGR} {\mathcal L} 
\newcommand{\MINK} {\sc minkowski}    
\newcommand{\Lagrangian} {{\sc lagrang\-}ian } 
\newcommand{\DEF} {\stackrel {def} = }

\newcommand{\PD}[2][]{\frac{\partial #1}{\partial #2}} 
\newcommand{\MAI} {\mathcal I} 

\title{ A new discrete view to quantum mechanics}
\author{Wolfgang Köhler, Potsdam, Germany}
\maketitle

\begin{abstract}
    Here I present  a new discrete model of quantum mechanics for relativistic 
    1-electron systems, in which particle movement is described by a directed
    space-time graph with attached 4-spinors, but without any continuous wave functions.
    These graphs only consist of few space-like edges, e.g. the ground state of atoms
    is described by two nodes and one edge,  and interactions only take place at the nodes.

    The fundament is an extremal principle for a relativistic invariant
    ``\Lagrangian sum'', from which ``field-equations'' and ``equations of motion''
    are derived, so the states (including the graph nodes) are completely determined.

    As important validations of the model,  the corresponding graphs for the stationary
    {\sc Dirac}-equation for the atom are drawn and the correct spectra are computed 
    ({\sc Sommerfeld}-levels). 
    
    Also a discrete {\sc schrödinger} approximation and an associated 
    ``{\sc hamilton}ian sum'' are derived and the correct equation of a classical 
    moving particle under {\sc Lorentz}-force is presented.
  
    I hope, that this new approach will help, to overcome some problems of current 
    quantum mechanics by making the wave function superfluous.
\end{abstract}
\tableofcontents

\section{Introduction}
In this paper, I propose a new discrete view to the quantum world, without the use of
a wave function concept. \\

The wave function was introduced by {\sc Erwin Schrödinger} around  1925
to describe quantum mechanical states, like electrons inside an atom, for which
classical descriptions failed.\\
 
However, there is a long, unceasing discussion about the interpretation
of this wave function, especially for the measurement process ("collapse" of wave function)
(see e.g. \cite{PENROSE-EMP} or \cite{BELL} pp. 40, for a comprehensive discussion).\\
On the other hand, it seems paradox, that the description of discrete
quantum states (like energy levels of an atom) required the invention of a new continuous
field. This new field has the additional strangeness, to be 'not physical', i.e. is not
directly measurable, like all other known fields.\\

Additionally, the current QED-theory has severe difficulties arising from infinite integrals,
which have to be eliminated by some mathematical tricks (renormalization). Many physicists believe,
that these are at least suspicious (see e.g. \cite{LL}, p. 458, \cite{BJORKEN}, pp. 166). 
As far as I see, the theory presented here,
will not show any infinite values.\\

As strong motivation to try discrete theories, I want to cite {\sc A. Einstein} 
from one of his last works (see \cite{EINSTEIN}, Appendix II, 1954,  p. 163):
 ``Man kann gute Argumente dafür anführen,
  daß die Realität überhaupt nicht durch ein kontinuierliches Feld dargestellt werden könne.
  Aus den Quantenphänomenen scheint nämlich hervorzugehen, daß ein endliches System von endlicher
  Energie durch eine  {\em endliche} Zahl von Zahlen (Quanten-Zahlen) {\em vollständig} beschrieben
  werden kann. Dies scheint zu einer Kontinuums-Theorie nicht zu passen und muß zu einem Versuch
  führen, die Realität durch eine rein algebraische Theorie zu beschreiben.
  Niemand sieht aber, wie die Basis einer solchen Theorie gewonnen werden könnte.''\FN{
  Emphasations by A. Einstein}\FN{
    Translation: 
    ``There are good arguments, that reality cannot be represented by a continuous field.
    It seems to follow from quantum phenomena, that a finite system of finite energy
    can be described {\em completely} by a {\em finite} number of numbers (quantum numbers).
    This seems not to fit to a continuum-theory and must lead to the attempt to describe
    reality by a pure algebraic theory. Nobody sees, however, how the basis for such
    a theory can be achieved. 
}\\

Nowadays, there exist several proposals to introduce 'discreteness' into physics.
Most of them postulate a space-time lattice at the {\sc Planck}-scale, of about $10^{-33} cm$ and
$10^{-44} s$. These scales are assumed to play a fundamental role in general relativistic 
quantum gravitation (which is not considered in this article).
However, due to the smallness of these units, it is not to expect to find consequences
of the lattice structure with currently available measurement techniques.
The quantum fields in these theories mostly appear as continuous approximations of discrete 
lattice fields.\\

My approach is different to the above mentioned, since it considers discreteness at particle
wavelength scales ($\sim 1/m$ for time-like edges, {\sc compton}-wavelength),\FN{
  In this article I use ``natural units'', where $\hbar = 1$ and $c=1$ holds.
} i.e. much larger, 
so it is directly related to the quantum nature of the particle.\\
On the other hand, I do not  describe the
whole space-time as a gridded structure, only the movement of the elementary particles
should be considered as not being continuous but in finite steps. There might exist
an underlying {\sc Planck}-scale grid, but this is not needed in the following 
considerations.\\

In this paper, I deal with a new view to special relativistic quantum mechanics of spin-1/2 particles 
in electromagnetic fields, i.e. {\sc Dirac} equation and their solutions 
(in flat {\sc minkowski} space-time), where the wave function is a 4-dimensional complex field.\\

In the discrete theory proposed here, the moving particle\FN{
  Of course, resting particles can be described as special stationary cases. 
} is described as a set of space-time points, with
finite space extent and, of course, infinite time extent. Attached to each time-like edge
is a {\sc Dirac}-spinor (as a constant) which replaces the continuous spinor field.\\

It turns out, that the number of points required to model
e.g. the energy levels of the atom, is only in the order of the quantum numbers.
In this (stationary) case the time-edges of the graph are simply equal and the space
edges  are constant. In nonstationary cases, however, also graphs with bifurcations 
and combinations are imaginable, but these are not considered in this article.\\

The equations for both grid-points and spinors are derived from an extremal
principle of {\em one general sum-function}. This seems to be the most appealing aspect
of this new theory. 
All other discrete theories, known to me, postulate some preset, fixed grids, which are not influenced 
by the fields.\FN{
  In some vague sense, 
  this resembles the concept of general relativity, where the space-time metric $g_{\mu\nu}$
  is influenced by the mass distribution.
}\\
This extremal principle resembles a {\sc Lagrange} functional, which is widely used in quantum physics
and esp. quantum field theory. 
Here, the \Lagrangian is replaced, of course, by a sum
over the space-time graph and the variational principle simply  maps to the variations
of the points and spinors.\\

To show the correctness of the theory two important cases are discussed and computed
below: the stationary electron in the {\sc Coulomb}-field (atom) and an electromagnetic 
acting particle in nonstationary case (accelerated by {\sc Lorentz}-force).\FN{
  Since most formulas derived there, are simple algebraic, they can easily be 
  implemented in numerical computer programs.
  In fact, I have done this for most of the examples, to validate the evaluations numerically.
}\\

A last word to the {\em structure} of this article. Some of the evaluations are not strictly
needed in this paper, esp. the simple start cases. However, it is to expect, that many readers
are not very familiar with the unusual notations used here. Thus, I think it is always better
to start with the simplest possible cases and then proceed to the more complex states.

In any case, I tried to put the evaluations straightforwardly as possible. 
So many proofs are left out, or shifted to the numerous footnotes, so that quick readers can skip them.
Longer ones were put into the appendix, which thus became quite voluminous.
It also serves, to illustrate the correspondencies to 
classical theories and formalisms.

\section{Notations, Entities and Transformations}
In contrast to the usual description of {\sc Dirac}-spinors as 4-spinors, I use a slightly
different notation by 'spinor-matrices'. These are complex 2x2-matrices, i.e. they
have the same number of components.\FN{
  They may be thought of writing the two bi-spinors as a two-column matrix, see Appendix. 
}\\
{\sc Minkowski}-vectors and {\sc lorentz}-transformations are then 
represented as certain {\em subsets} of these 2x2-matrices, with constraints explained below.

In the appendix it is shown, that both notations are equivalent
for the {\sc Dirac}-equation and classical relativistic electromagnetism.\\

The main reason for
using this form is, that {\em all entities} are represented by the {\em same algebraic structure}, and
many of the following equations are much better readable, than in component notation.\\

General 2x2-matrices are here denoted with uppercase letters $P,Q,S,\dots$.\\
The usual operations with the matrix $P = {a,b\choose c,d}$, with complex $a,b,c,d$, 
here are written as:\FN{
$I$ denotes the 2x2-identity matrix.}
\begin{itemize}
\item $\bar P = {\;\; d, -b\choose -c,\;\; a}$ : adjuncted matrix of $P$,\FN{
  The inverse matrix of $P$ is then, of course $P^{-1} = \bar P/|P|$.
}
\item $|P| = ad-bc$ : the scalar determinant, with  $P\bar P = I\;|P|$,
\item $\TR P =  a  +d$: the scalar trace, with $P + \bar P =  I\; \TR P $,
\item $P^T = {a,c\choose b,d}$ : the transposed, 
  $P^\ast = {\CC a,\CC b\choose \CC c,\CC d}$ the complex conjugated,
\item $\HC P = (P^\ast)^T = {a^\ast, c^\ast \choose b^\ast , d^\ast}$  : 
  the adjungated (or {\sc hermite}an conjugated).\FN{All operations here 
    commute, e.g. $\BAR{(\HC P)} = \HC{(\bar P)}$ and products obey 
  $\HC{(PQ)} = \HC Q \HC P$ and $\BAR{(PQ)} = \bar Q \bar P$.
}
\end{itemize}
  
Since it is often needed in the following, and not quite obvious, I state 
here  the general circularity relation for the trace of any matrices $ABC,\dots, X$:
$\TR{A\; BC\dots X} = \TR{BC\dots X\; A} $.\FN{It can be derived from the symmetry relation 
  $\TR{AB} =\TR{BA}$, which again follows e.g. from 
  $|A+\bar B| = (A+\bar B)(\bar A + B) = |A| + |B| + \TR{A B} = | B + \bar A|$.
}\\

A {\sc Minkowski}-{\em vector} is
in this formalism represented by a {\sc hermite}an matrix, and here denoted by boldface (upper-
and lowercase) letters:\FN{
  One exception is the relativistic $\partial$-operator.
} $\HC {\MV M} = \MV M$, to distinguish it from other matrices (spinors, transformations, electromagnetic
field tensor).\\
It has, of course, 4 real components, which can be mapped
to space-time coordinates $(t,x,y,z)$ in the following way (see e.g. \cite{PENROSE}, pp. 16):\FN{  
  Since every matrix $M$ can be uniquely decomposed into a {\sc hermite}an and anti-{\sc hermite}an part  
  by $M = \MV A + i \MV B$, with $ \HC {\MV A} = \MV A,\; \HC {\MV B} = \MV B$
  (which then transform independently under {\sc lorentz}-{transformations}),
  all 2x2 matrices can be seen as generalization of {\sc Minkowski}-matrices.
}
\begin{equation}
  \label{eq-1} 
  \MV M = { t+z, x-iy\choose x+iy,t-z} = t I + x\sigma_1 + y\sigma_2 + z\sigma_3,
\end{equation}
where $\sigma_i$ are the usual {\sc Pauli}-matrices.\\

{\sc lorentz}-{\em transformations} are represented by unimodular matrices $T, |T| = 1$, and thus
have 6 real degrees of freedom.\FN{
  Mathematically speaking, in terms of {\sc Lie}-group theory, $T$ build the $SL(2,\mathcal C)$-group,
  which is a double cover of the {\sc lorentz}-group.
}
Ordinary space rotations additionally fulfill the condition
$\HC T = \BAR T$, leaving 3 free real degrees,\FN{
  By this definition they build a subgroup (the {\em quaternion group}, see appendix), 
  while special {\sc lorentz}-transformations do not.
} while special {\sc lorentz}-transformations obey $\HC T = T$.\FN{E.g. a  matrix 
$T = {\beta,0 \choose 0,1/\beta}$ with
real $\beta$, performs a $(t,z)$ transformation.
}\\

With respect to their behaviour under space-time transformations, we must distinguish between 
 spinor- and {\sc Minkowski}-matrices. 

Let $T$ be a {\sc lorentz}-transformation, then a spinor transforms
with $P \to T P$ (then follows e.g. $\HC P \to \HC P \HC T$), while
a {\sc Minkowski} matrix transforms with $\MV M \to T \MV M \HC T$.\FN{
  The above condition for space rotations $\HC T = \BAR T = T^{-1}$ then leads to  $\MV M \to T \MV M T^{-1}$, 
  consequently the trace of ${\TR {\MV M} = 2t}$ is invariant, as required.    

  Consider e.g. the transformation $T = i\sigma_1$, which performs a rotation  of 180° around the $x$-axis.
  A full rotation is then represented by $T = (i\sigma_1)^2 = -1$.
}\\

The determinant $|\MV M|$ is then obviously the {\sc Minkowkski}an {\em invariant}\FN{
  The proof of invariance is simple: $ |\MV M| \to |T| |\MV M| |\HC T| = |\MV M|$, since $|T|=1$
} (always real):
\begin{equation}
  \label{eq-2} |\MV M| = t^2 - x^2 - y^2 -z^2.
\end{equation}
It is remarkable in this formula, that the signature of the metric tensor ($+$$-$$-$$-$) automatically 
follows from the property of {\sc hermite}city.

To build the general {\em scalar product} of two {\sc Minkowski} matrices $\MV A, \MV B$ serves
the formula, which is  obviously also invariant and real:
\begin{equation}
  \label{eq-3} \TR{\MV A \bar {\MV B}} = \TR{\MV B \bar {\MV A}}.
\end{equation}

From the representation (\ref{eq-1}) should also be noted, that the trace of a {\sc minkowski} 
matrix maps to the {\em time component}, and the operation of ``adjunction'' is a space ($R^3$) inversion.

\section{Space-Time Graph and ``{\sc Lagrange}-Sum''}

\subsection{Descriptions of Particle Movement in SRT}\label{sec-srt}
This short section is intended to explain concisely, how particle movement is described in the 
context of {\em Special Relativity} and {\sc minkowski} space-time. 
Also a discrete variant of movement (which is not used in the following) is sketched, 
but quantum effects are not considered here.\\

A {\em continuous relativistic particle trajectory} is given by a space-time curve i.e. the 4 functions
$\MV x(\tau) = (x(\tau), y(\tau), z(\tau),t(\tau))$\FN{
  Or in matrix notation written as:
  $\MV x(\tau) = {t(\tau) + z(\tau), x(\tau) - i y(\tau) \choose x(\tau) + i y(\tau), t(\tau) - z(\tau) }$.
}
which is usually parametrized by the {\em eigentime} $\tau$.\FN{
  But also other parameters, e.g. $t$ may be used. The advantage of using $\tau$ is, that the
  velocity vector $\MV u = \frac {d\MV x}{d\tau }$ is then normalized to unity.
}\\
This $\tau$ is defined by $d\tau^2 = |d\MV x| = dt^2 - dx^2 -dy^2 - dz^2$.\\

Since $d\tau^2$ is an invariant, all {\sc minkowski}-vectors $d\MV x$ can be classified by its sign:
$d\tau^2 > 0$ : time-like, $d\tau^2 = 0$ : light-like, $d\tau^2 < 0$ : space-like.\\

Usually by the condition of causality it is required, that no interactions over space-like separated
regions occur.
The movement of a particle  is restricted to time-like vectors 
(resp. light-like for massless particles).\\

The {\em discrete} form of such a space-time curve is then simply a sequence 
\[\MV x_1 = (x_1,y_1,z_1,t_1),\quad \MV x_2 = (x_2,y_2,z_2,t_2),\quad \dots \]

This sequence can be considered as a graph with the edges $(\MV x_k \to \MV x_{k+1})$,
which describe a movement in finite ``jumps'' and ``time-likeness'' 
here obviously means $|\MV x_{k+1} - \MV x_k| > 0$.\\

The (continuous) movement of a classical charged particle in an electromagnetic 
field ({\sc Lorentz}-force) can also be
derived from a  variation principle for a \Lagrangian.

Let all possible space-time curves be parametrized by a parameter $\lambda$ : $\MV x(\lambda)$.
In matrix notation the action integral is then written:\FN{
  absolute value of a {\sc minkowski} vector written as 
  $|| \MV x|| \stackrel {def}= \sqrt{ |\MV x| }$
} 
\[ 
\LAGR(\MV x(\lambda)) = \int\limits_{\lambda_1}^{\lambda_2} d\lambda 
\left[ m \left|\left|\frac {d\MV x}{d\lambda} \right|\right| + 
  e \TR{\bar \MV A(\MV x) \frac {d\MV x}{d\lambda}} \right].
\]
The variation of space-time curve
$ \MV x(\lambda) = \MV x_e(\lambda) + \delta \MV x(\lambda) $
results in an extremal curve $\MV x_e(\lambda)$.\\ 
Then $\lambda$ is identified with the eigentime of the extremal curve  $d\tau \stackrel {def}= ||d \MV x_e||$.\\

The extremal curve is then given by the equation of {\sc Lorentz}-force:\FN{
  See Appendix for the relations between $\MV A$ and $F$ in matrix notation.
}
\[ m \frac {d^2\MV x}{d\tau^2} = 
\frac e2 \left( \frac {d\MV x}{d\tau} F +  \HC F \frac {d\MV x}{d\tau} \right).
\]

It is interesting to state here, that there exists also {\em discrete representations} of the above
\Lagrangian formalism. 
However, this example is given as illustration only, and {\em not used in the following
sections!}

A possible {\em discrete} variant of the above integral is:
\[ 
\LAGR = m \sum_k ||\MV x_{k+1} - \MV x_k || + e \sum_k \TR{\bar \MV A (\MV x_k)(\MV x_{k+1} - \MV x_{k-1})}.
\]
Herein the variation of one space-time point (node) $\MV x_k$ 
leads to a {\em discrete} version of the {\sc Lorentz}-Force:
\[ m\left(\frac {\MV x_{k+1} - \MV x_k}{||\MV x_{k+1} - \MV x_k||}
-\frac {\MV x_k - \MV x_{k-1}}{||\MV x_k - \MV x_{k-1}||} \right)
\approx \frac e2 \left((\MV x_{k+1} - \MV x_{k-1})F_k +  \HC F_k (\MV x_{k+1} - \MV x_{k-1})\right).
\]
Like in the following sections, here also the identifaction $||\MV x_{k+1} - \MV x_k|| = 1/m$ is 
possible, since this expression is conserved (approximately).\FN{
  This has no analogon in the continuous case.
  To prove it, multiply the eq. with $\bar \MV q \stackrel {def}= \bar \MV x_{k+1} - \bar \MV x_{k-1}$ 
  (e.g. from left) and take the trace (i.e. building scalar product with $\MV q$),
  Then the r.h.s. vanishes, since $\TR{F} = \TR{ \HC F} = 0$.
}

\subsection{General Considerations with Space-Time Graphs}
Space-time grids are commonly used to solve partial differential equations,
e.g. numerically. Then integrals (e.g. the \Lagrangian functional) are
represented as sums. Usually, the gridded structure is viewed as approximation
of the continuum, and the smaller the edges are, the better the
approximation.\\
In this theory, I try another point of view: the grid  represents the quantum state and the differential 
form is the approximation.\\
In fact, it turns out in the following, that e.g. to describe bound states in the atom, that
there exist ``minimal grids'', which suffice to represent the {\em exact states} in the 
{\sc  dirac}-theory.\\
It should be noted, however, that the usage of {\em finite} (esp. space-like) edges introduces some
kind of nonlocality and causality violation into the theory.\FN{
  Also should be added, that the used graphs show some
  similarities to {\sc feynman}-graphs. However, bifurcations are not considered here, and
  the mathematical background is completely different.
}\\

I will start with the general expression for the 
``\Lagrangian sum'' over any graph. Let $\{\MV x_i \}$ be the nodes as {\MINK} matrices 
(numbered in an arbitrary order), $\MV H_{ij}, \MV M_j$ be some {\sc Minkowski}an matrices,  
 $\MV A$ the electromagnetic vector potential, which
has the values $\MV A_{i} \stackrel{def}= \MV A(\MV x_{i})$ at the grid points, and $P_i$ some 
spinors:\FN{
  As usual, $m$  denoting particle mass,
  $e$ electrical charge and $\Re()$ the real part of a complex number.
}

\begin{equation}
\label{eq-4}
\LAGR = \sum_{ij}\TR{ (\MV x_{i} -\MV x_{j})^{-1} \MV H_{ij}} 
+ e\sum_{i}\TR{\bar \MV A_{i} \MV M_{i}} -   2m \sum_{i} \Re(|P_{i}|).
\end{equation}
However, the first double sum is not to be applied for all pairs $(i,j)$, but only for edges.\\
The auxiliary matrices $\MV H_{ij},\MV M_i$ in this equation shall be constructed as {\sc hermite}an 
bilinear forms from the fundamental spinors $P_i$, as explained below.\\
 
At first, however, it is to prove, that this sum fulfills all requirements for a \hfill\\ 
``{{\sc Lagrang}ian}'': it is a 
real scalar and invariant under all {\sc lorentz} transformations.\\ 

It is obviously scalar, by construction. To  prove the reality of $\LAGR$, I state  that:
\begin{itemize}
\item for any  {\sc hermite}an matrix $\MV A$ holds trivially: $\TR {\MV A} = real$,

\item for any two {\sc hermite}an matrices $\MV A,\MV B$ holds: $\TR{\MV A\MV B } = real $,
  due to the symmetry relations:
$\TR{\MV {A B} } = \TR {\MV{BA} } = \TR{\MV{\HC B\HC A}} = \TR{\HC{(\MV  {AB})}} $.

\end{itemize}

Since all factors are {\sc hermite}an matrices and $\Re(|P|)$ is always real, the complete
sum is real.

To prove {\sc lorentz} invariance, I state, that the epressions ${\MV x}^{-1}$ and $\MV {\bar A}$
transform with $\bar\HC T (\cdot) \bar T $ and therefore the expressions $\TR{\MV x^{-1} \MV H}$ and
$\TR{\bar \MV A \MV M}$ are invariant scalar products.\\ 
The determinant $|P|$ is trivially invariant under {\sc lorentz}-transformations, if the
spinor transformation rule  $P\to T P$ is considered, q.e.d.\\

\subsection{Regular Space-Time Graphs}

Now I consider {\em regular} space-time graphs. The restriction to these graphs is mainly due to
the problem, that it is not yet clear, what physical conditions can lead to bifurcations or
combinations, and the mathematical difficulties in handling them.\\
This is no principal limitation,
and as shown in the following, many problems of one-particle quantum mechanics can be described with
these graphs.\\

One first introduces double indices for the nodes
$\{\MV x_{ik}\}$, where
the first index should stand for space, while the second index 
$k$ stands for time steps (thus unbounded, $ k = -\infty \cdots \infty$).\FN{
  This numbering scheme does not violate the {\sc lorentz}-covariance of
  the following evaluations.
}

The regularity condition then means, for any time index $k$
there exist $n$ nodes: $i = 1,\dots, n$  and that for any $i,j,k$ should
hold $|\MV x_{ik} - \MV x_{jk}| < 0$ (space-like edge), 
and for any $i,k$:  $|\MV x_{i,k+1} - \MV x_{i,k}| > 0$ (time-like edge)
and also $t_{i,k+1} - t_{i,k} = \frac 12 \TR{\MV x_{i,k+1} - \MV x_{i,k}} >0$ 
(direction of time arrow). 
Also, only  timely consecutive nodes $(k, k+1)$ shall be connected by an edge.\\

The (constant) spinors $P_i$ are now considered to be assigned uniquely to the {\em time-like edges} 
(not to the nodes): $ (\MV x_{i,k+1},\MV x_{i,k}) \leftrightarrow P_{ik}$.\\
Of course, this assumption introduces a fundamental asymmetry between space and time and
leads to different formulas for the above introduced $\MV H$.\\
For space-symmetry reasons, the following ansatz is suggested:\\  
$\MV H_{i,k+1,i,k} = P_{ik} \HC P_{ik},\quad \MV H_{i,k,j,k} = P_{i,k-1} \circ P_{j,k}$,\FN{
The circle stands for the {\sc hermite}an conjugated expression:
$P\circ Q \stackrel {def}=\frac 12 (P\HC Q + Q \HC P)$.
} (all other combinations of indices have no edge assigned)
and
$\MV M_{ik} = \frac 12 (P_{ik} \HC P_{ik} + P_{ik-1} \HC P_{ik-1})$.\FN{all obviously {\sc hermite}an
}\\
Then from (\ref{eq-4}) results:
\begin{eqnarray}
\label{eq-5}
\LAGR &=& \sum_{k,i} \TR{(\MV x_{i,k+1} -\MV x_{i,k})^{-1}P_{ik}\HC P_{ik} }\nonumber\\
  &+& \sum_{k,ij}\TR{(\MV x_{i,k}-\MV x_{j,k})^{-1}\frac 12(P_{i,k-1}\HC P_{j,k}+P_{j,k}\HC P_{i,k-1})}\\
  &+& e\sum_{k,i}\TR{ \frac 12(\MV {\bar A_{i,k+1}} + \MV {\bar A_{i,k}}) P_{ik}\HC P_{ik}} 
 - 2 m \sum_{ik} \Re(|P_{ik}|). \nonumber 
\end{eqnarray}

To visualize the kinematic terms (first and second term) of this sum, the following picture is used, 
where the spatial extent number is set to $n=2$ (this example graph e.g. also represents the ground state
of an electron in an atom):\\

\setlength{\unitlength}{1cm}
\begin{picture}(10,5.0)
\label{fig1}
\multiput(3,1.1)(0,1.0){4}{\vector(1,0){4}}  
\multiput(7,1.1)(0,1.0){4}{\vector(-1,0){4}}  
\multiput(3,1.1)(0,1.0){4}{\circle*{0.1}}  
\multiput(7,1.1)(0,1.0){4}{\circle*{0.1}}  

\multiput(3,0.1)(0,1.0){5}{\vector(0,1){1}}    
\multiput(7,0.1)(0,1.0){5}{\vector(0,1){1}}    

\newcounter{mycount}
\setcounter{mycount}{1}
\multiput(1.5,1.0)(0,1.0){4}{
  {$ \MV x_{1\arabic{mycount}}, \MV A_{1\arabic{mycount}}$}\stepcounter{mycount}
}
\setcounter{mycount}{1}
\multiput(7.2,1.0)(0,1.0){4}{
  {$\MV x_{2\arabic{mycount}}, \MV A_{2\arabic{mycount}}$}\stepcounter{mycount}
}

\newcounter{ic1}
\newcounter{ic2}
\multiput(4,1.2)(0,1.0){4}{ 
\stepcounter{ic2} $ P_{1\arabic{ic1}} \circ P_{2\arabic{ic2}}$ \stepcounter{ic1}
}
\setcounter{ic1}{0}\setcounter{ic2}{0}
\multiput(4,0.8)(0,1.0){4}{ 
\stepcounter{ic2} $ P_{1\arabic{ic2}} \circ P_{2\arabic{ic1}}$ \stepcounter{ic1}
}
\setcounter{ic1}{0}
\multiput(2.2,0.5)(0,1.0){5}{$ P_{1\arabic{ic1}}$ \stepcounter{ic1}}
\setcounter{ic1}{0}
\multiput(7.3, 0.5)(0,1.0){5}{$ P_{2\arabic{ic1}}$ \stepcounter{ic1}}
\end{picture}\\[0.05cm]
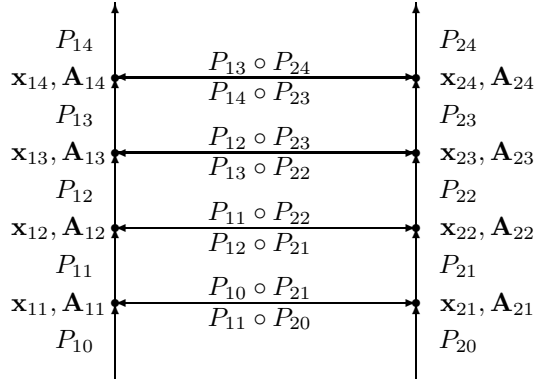
\centerline{{\bf Fig. 1: example space time graph} (time axis vertical)}\\

The {\em extremal principle} now considers the sum $\LAGR$ as a function of all inner variables 
$\MV x_{ik}, P_{ik}$, 
whereas possibly some boundary variables have to be fixed, to account for initial conditions:
\begin{equation}
  \label{eq-6} 
  \LAGR(\MV x_{ik}, P_{ik}) \to Extr.
\end{equation}

\subsection{Simplest Case} 
To demonstrate the method of deriving the ``field equations'' and ``equations of motion''
 from this principle, I start with
the simplest case: no electromagnetic potential ($\MV A = 0$) and 
the graph has only one spatial index: $n = 1$.\\
This model represents a freely moving spin-1/2 particle.

The graph then reduces to a sequence of {\sc minkowski} space-time points, 
which is actually a discrete particle trajectory, as explained in the section \ref {sec-srt}.

The spatial index can be omitted, and the second and third sum in equation (\ref {eq-5}) are zero.
It remains the sum:
\begin{equation}
  \label{eq-7}
  \LAGR (\MV x_k, P_k) =  \sum_{k} \TR{(\MV x_{k+1} -\MV x_{k})^{-1}P_{k}\HC P_{k} }
  - 2 m \sum_{k} \Re(|P_{k}|).
\end{equation}
This sum is obviously invariant under the "local" 
transformations $P_k \to P_k S_k$, when $S_k\HC S_k = |S_k| = 1$.\FN{
  These $S$ are
  normalized {\em quaternions}, see appendix.
}
That means, the spinors $P_k$ are determined only up to these factors by the following
equations (gauge invariance).\\ 

At first, I consider the variation of one specific $P_k$, in this sum. This variation
is similar to the usual methods in quantum theories.\\
For the variation of the
real part of the determinant, is used: $2\Re(|P|) = |P| + |\HC P|$ and 
\begin{equation}
  \label{eq-8}
  |P +\VAR P| = (P +\VAR P)(\BAR{ P+\VAR P}) \approx |P| + \VAR P \BAR P + P \BAR{\VAR P}
  = |P| + \TR{\VAR P \BAR P},
\end{equation}
 therefore results from substitution $P_k \to P_k +\VAR P_k$ the variation
\begin{equation}
  \label{eq-9} 
  \VAR \LAGR =  \TR{(\MV x_{k+1} -\MV x_{k})^{-1}(\VAR P_{k} \HC P_{k} + P_k\VAR \HC P_k }
  - m  \TR{\VAR P_{k} \BAR P_{k} + \VAR \HC P_k\VAR \BAR{\HC P_k}}.
\end{equation}
To simplify the formulas, I define an auxiliary variable 
$\MV v_k \stackrel {def}= (\MV x_{k+1} -\MV x_{k})^{-1}$, so equation (\ref {eq-9}) writes
(after using the general circularity relations for the trace):
\begin{equation}
  \label{eq-10}
  \VAR \LAGR =  \TR{\VAR P_{k}(\HC P_{k}\MV v_k - m  \BAR P_{k})}
  +\TR{\VAR \HC P_k(\MV v_k P_k - m \BAR{\HC P_k})}.
\end{equation}
As usual, the variations of $\VAR P_k$ and $\VAR \HC P_k$ are considered as independent,
therefore both terms must vanish. An expression $\TR{XY}$, however, can only vanish for
any matrix $X$, if $Y = 0$ holds. So the two (equivalent, since by 
definition $\MV v = \HC {\MV v}$) equations result:
\begin{equation}
  \label{eq-11}
  \HC P_{k}\MV v_k =  m  \BAR P_{k}\MBOX{and}\MV v_k P_k = m \BAR{\HC P_k}.
\end{equation}
This equation corresponds to the usual ``field equations'' in quantum mechanics.
Here it forces (by taking the determinant on both sides) that, 
since $|\MV v_k| = real$, also $| P_k| = |\HC P_k| = real$, 
consequently $|\MV v_k| = m^2 = const.$ or $|\MV x_{k+1} -\MV x_{k}| = 1/m^2$, 
implying that the motion vector is a time-like vector of the constant length $1/m$.\FN{
  In the rest frame of the particle, thus trivially holds $\Delta t = 1/m$, for all others
  relativistically   $\Delta t > 1/m$.
}\\
  
The {\em second variation}\FN{
  This is not to misinterprete as a {\em  second order} variation.
} (which has no correspondence in current theories) varies the nodes $\MV x_k$.
For this, it is needed to state, that for small $\VAR \MV x << \MV x$ holds (see appendix)\FN{
  A simple proof is, to multiply the equation from the left (or right) with $(\MV x + \VAR \MV x)$.
}
\begin{equation}
  \label{eq-12} 
  (\MV x + \VAR \MV x)^{-1} \approx \MV x^{-1} -\MV x^{-1}\VAR\MV x \MV x^{-1}.
\end{equation}
The variation of $\MV x_k$ influences only $\MV v_k$ and $\MV v_{k-1}$ (by their definition)
and leads to
\begin{equation}
  \label{eq-13}
  \VAR \MV v_k =  +\MV v_k\VAR\MV x_k \MV v_k\MBOX{and}
  \VAR \MV v_{k-1} =  -\MV v_{k-1}\VAR\MV x_k \MV v_{k-1}
\end{equation}
The variation of $\LAGR$ in (\ref{eq-7}) is consequently (second line by circulation):
\begin{eqnarray}
  \label{eq-14}
  \VAR \LAGR &=& \TR{\MV v_k\VAR\MV x_k \MV v_kP_k\HC P_k} -
  \TR{\MV v_{k-1}\VAR\MV x_k \MV v_{k-1}P_{k-1}\HC P_{k-1}}\nonumber\\
  &=& \TR{\VAR\MV x_k(\MV v_kP_k\HC P_k\MV v_k - 
    \MV v_{k-1}P_{k-1}\HC P_{k-1}\MV v_{k-1})}.
\end{eqnarray}
Again, this expression must vanish for arbitrary $\VAR \MV x_k$, leading to
\begin{equation}
  \label{eq-15}
  \MV v_kP_k\HC P_k\MV v_k \stackrel !=
  \MV v_{k-1}P_{k-1}\HC P_{k-1}\MV v_{k-1},
\end{equation}
Inserting equations (\ref{eq-11}) twice, results in $ P_k\HC P_k =  P_{k-1}\HC P_{k-1} = const.$.
This is fulfilled by the condition $P_k = P_{k-1}S$, with an arbitrary matrix,
that obeys $S\HC S= I$ (gauge invariance).\\
On the other hand, follows from equation (\ref{eq-15}):\FN{
  The other solution 
  $ \MV v_k = -\MV v_{k-1}$ would imply a step backwards in time.
}
\begin{equation}
  \label{eq-16} \MV v_k = \MV v_{k-1},
\end{equation}
that says, that the motion vector is constant over time, as it should be.\\

The result of above computations is, that the particle ``moves'' in jumps with a constant
time-space vector $\MV x_{k+1} - \MV x_k = \Delta \MV x$, which is related to the particle
wavelength by $\sqrt{|\Delta \MV x|} = 1/m$. The {\sc Minkowski} vector $\BAR {\MV v_k} = const.$ is
therefore to identify with the relativistic energy-impuls vector 
$\MV p = \varepsilon + \vec p $, where $\varepsilon$ denotes the energy and it holds
$|\MV p| = \varepsilon^2 - |\vec p|^2 = m^2 $.\\
The spinor orientation does not have any influence on the motion, as to expect in the 
absence of an external field.\\

In any case, the time-steps that arise in these jumps are by orders too small to be
visible in experiments (e.g. with ultrashort laser pulses). For an electron e.g. holds
$\Delta t \approx 10^{-20} s$.

\subsection{Stationary Case}
This case describes bound states, e.g. an electron in an atom. 
I will show in the following, that it leads to the correct energy spectrum.\\

The grid for this case is considered as ``time invariant'',
i.e. it is claimed $\MV x_{i,k+1} - \MV x_{ik} = \tau$, 
where $\tau = real,\; \tau > 0$ is a constant time step.\FN{
  Like above, it can be identified with the
  inverse energy of the state and its value follows from the equations below as
  the {\em eigenvalue}.
}\\ 
Only in this ``periodic'' case, the space components of the graph are constant over time (for all
time indecees).\\
The space-like edges shall be pure space vectors (traceless matrices 
${\bar {\MV x} = -\MV x }$).\\

 Additionally, for all $i,k$ the ansatz $P_{i,k+1} = P_{i,k} S$ is made\FN{
   Considering $S$ beeing independent of the spatial index $i$ is the standard
   method of ``separating variables'' (here time and space are to separate). 
   In function form one would make e.g. the ansatz $\Psi(x,t) =\phi(x)\psi(t)$ 
   and here $S^k$ stands for the time dependency factor.
} 
(i.e. $P_{i,k} = P_{i,1}S^{k-1}$)
with a constant matrix $S$, obeying $S\HC S = I$ and $|S| = 1$.\FN{
  These conditions hold for all {\em unit quaternions} $S$, which is equivalent to $S\in SU(2)$.
}\\

It follows $|P_{i,k}| = | P_{1,k}|, \quad 
P_{i,k+1} \HC P_{i,k+1} =P_{i,k} \HC P_{i,k} =\cdots = P_{i,1} \HC P_{i,1} = const.$ and
 $ P_{i,k-1} \HC P_{j,k} =P_{i,k-1} \HC S \HC P_{j,k-1} = \cdots = P_{i,1} \HC S \HC P_{j,1}$.\\

The summands of $\LAGR$ in (\ref{eq-5}) then become independent of the time index $k$, if
also the external field is considered as time invariant: $\MV A_{i,k} = \MV A_{i,1}$. 
The index $k=1$ can be dropped, also the summation over $k$ can be omitted and one gets: 
\begin{eqnarray}
  \label{eq-17}
  \LAGR &=& \sum_i \TR{(\frac 1\tau + e\bar \MV A_i) P_i\HC P_i} +
  \frac 12\sum_{ij}\TR{(\MV x_i -\MV x_j)^{-1}(P_i\HC S \HC P_j+ P_j S \HC P_i)}\nonumber \\ 
&-& 2m \sum_i\Re(|P_i|).
\end{eqnarray}
Again, the double sum is to build only over edge-pairs $(i,j)$.\\

To simplify the formulas, I introduce a set of auxiliary variables
\[ \MV u_{ij} \stackrel {def}= (\MV x_i -\MV x_j)^{-1} = - \MV u_{ji}.
\]
The antisymmetry of the factor $\MV u_{ij}$ in the double sum in $i,j$ leads to 
a simplification, e.g. the summands for the pair $(1,2)$  are:
\[
  P_1\HC S \HC P_2+ P_2 S \HC P_1 - P_2\HC S \HC P_1- P_1 S \HC P_2
  = P_1(\HC S - S) \HC P_2+ P_2 (S -\HC S) \HC P_1.
\]
 Any unit quaternion $S$ can generally be represented
 with real $\lambda$ and $U$ as general, pure vectorial, unit quaternion 
 ($\HC U = \bar U = -U$, $|U| = 1$)\FN{
   $U$ has 2 free real parameters and its general form is 
   $U={i \sin\varphi, \;\;\; \cos\varphi e^{i\chi} \choose \cos\varphi e^{-i\chi}, -i \sin\varphi }$.
 } as:
\begin{equation}
  \label{eq-18}
  S = e^{\lambda U} = \cos\lambda + U\sin \lambda.
\end{equation}
This gives $S-\HC S = 2U\sin\lambda$. 
Inserting this  in (\ref{eq-17}) follows, that the double sum is proportional to $\sin\lambda$ (which
is the only term containing $\lambda$). 
The extremal principle then requires (since $\lambda$ is a free ansatz-parameter), that 
$\cos\lambda \stackrel != 0$, therefore $S = \pm U$ and $\HC S = - S$,\FN{
  In the following one may use e.g. $S = U = {i,\;\; 0\choose 0,-i}$, the explicit form does not matter.
}
and equation (\ref{eq-17}) simplifies to (here also $\tau$ is replaced by the energy $\varepsilon = 1/\tau$)
\begin{equation}
  \label{eq-19}
  \LAGR = \sum_i \TR{(\varepsilon + e\bar \MV A_i) P_i\HC P_i} +
  \sum_{i j} \TR{ \MV u_{ij}P_j S \HC P_i} -2m \sum_i\Re(|P_i|).
\end{equation}
The variation of $P_i$ is carried out like in the last section and leads to the
``field equation'' (a system of $n$ linear equations for the $P_i,\; i=1,\dots,n$):\FN{
Note, that the sum in this equation, in contrast to the sums above, is simple, since $i$ is fixed.
}
\begin{equation}
  \label{eq-20}
  (\varepsilon + e\bar \MV A_i)P_i + \sum_{ij}\MV u_{ij}P_j S = m \bar \HC P_i
\end{equation}

This equation becomes equal to the {\sc Dirac}-equation for the stationary case, if
the ``operator'' $\sum_{i j}\MV u_{ij}$ is replaced by the spatial derivative 
($\nabla$-operator). 
This correspondence is shown in the appendix.\FN{
  If one consideres this equation as a classical
  eigenvalue-problem for $\varepsilon$ (given $\MV x_{i}$ and $\MV A_i$) it has, of course,
  at least $n$ solutions $\varepsilon_i$ 
  and corresponding eigenvectors (at most $4n$, because every matrix-equation has actually
  4 scalar equations).
}\\

Again, I consider a {\em second variation} of the grid points $\MV x_i$. 
The  ``equations of motion'' derived with
this method, result in the determination of the grid points. 
Again, this procedure has no counterpart in the present theories.\\
Please consider again, that this variation does {\em not affect} the spinors $P_i$.\\
In the next section I will show, that it produces the correct quantum states for the
electron in {\sc Coulomb}-potential.\\

For the variation of $\MV x_i$ in the sum (\ref{eq-19}), at first it is needed, that\FN{
see appendix for a short explanation.}
\begin{equation}
  \label{eq-21}
  \VAR \MV A_i = \VAR \MV A(\MV x_i) = \frac 12 \TR{\VAR \MV x_i \bar \partial} \MV A_i,
\end{equation}
where $\partial$ is the {\sc minkowski}an differential operator (an explicit representation in
matrix notation is given in the appendix).

The variation of $\MV u_{ij} = (\MV x_i -\MV x_j)^{-1}$ is again
\begin{equation}
  \label{eq-22}
  \VAR \MV u_{ij} = -  \MV u_{ij}\VAR \MV x_i \MV u_{ij} \MBOX{and}
  \VAR \MV u_{ji} = + \MV u_{ji}\VAR \MV x_i \MV u_{ji}.
\end{equation}
 It results for the variation of $\MV x_i$, when another auxiliary variable 
$H_{ij} \stackrel {def}= P_i S \HC P_j$ (with $\HC H_{ij} = -H_{ji}) $ is
introduced for simplification (note, that in the sum $i$ is fixed)
\FN{
the differential operator here operates only
on the external field $\MV A$, of course, since $P_i$ is considered as a constant.} 
\begin{equation}
  \label{eq-23}
  \VAR \LAGR = \frac e2\TR{\VAR \MV x_i \BAR \partial}\TR{ \MV A_i P_i \HC P_i} + 
  \sum_{ij} \TR{ \MV u_{ji}\VAR \MV x_i\MV u_{ji}H_{ij} -
    \MV u_{ij}\VAR \MV x_i\MV u_{ij}H_{ji}}.
\end{equation}
From the demand $\VAR \LAGR = 0$ for all $\VAR \MV x_i$ (with operations like in previous
section), results the equation\FN{for the important special case, that $\MV A$ is time constant, 
  $(\partial +\bar\partial) \MV A = 0$ holds and therefore from (\ref{eq-24}) can be derived 
  $\sum_{ij} \MV u_{ji}^2\TR{H_{ij} + \HC H_{ij}} = 0$, since $\MV u_{ji}^2 = scalar$.
}
\begin{equation}
  \label{eq-24}
  \frac e2\BAR \partial\TR{ \MV A_i P_i \HC P_i} + 
  \sum_{ij}\MV u_{ji}(H_{ij} + \HC H_{ij})\MV u_{ji} = 0.
\end{equation}
The combined solution of (\ref {eq-20}) and (\ref {eq-24}) is then the expected quantum
state, which also determines $\varepsilon$, as it is to see in the following examples.\\

However, sometimes it is easier to use another method, that obviously leads to the same
results. For all solutions of equation (\ref{eq-20}) follows, that $\LAGR = 0$  holds. 
Then $\LAGR(\varepsilon,p_1,p_2,\dots) = 0$ is an implicit equation 
for $\varepsilon$ ($p_i$ subsumming all free variables, here as simple reals). 
Since $\LAGR$ shall be extremal with respect to all
other parameters $p_i$ it follows $\displaystyle \PD[\varepsilon]{p_i} = 0$, 
so $\varepsilon(p_1,\dots)$ itself must be extremal (usually minimal).\\

\section{Electron in an Atom}
In the usual approximation, the atom  nucleus shows a {\sc coulomb} potential, 
leading to scalar $e \MV A = eV(r) = \frac\alpha r$, where $\alpha = e^2 \approx 1/137$ 
denotes the ``finestructure constant'' and $r = \sqrt {x^2+y^2+z^2}$ the 
{\sc euclid}ian distance to the center.\FN{In matrix notation holds $r = \sqrt{-|\MV x|}$.} 

\subsection{{\sc Coulomb}-potential}
\label{sec-coulomb}
To solve the equation (\ref{eq-20}) we first rewrite it for {\sc coulomb}-potential,
by using the auxiliary parameters $\varepsilon_j \stackrel {def}= \varepsilon + \alpha/r_j$, and the same
with conjugation\FN{consider $\BAR {\MV u} = - \MV u $, $\HC {\MV u} = \MV u$ and
$\BAR{\HC S} = S$.}
\[
  \varepsilon_i P_i + \sum_{i\ne j}\MV u_{ij}P_j S = m \BAR{\HC P_i}\MBOX{equals}
  \varepsilon_i \BAR{\HC P_i} - \sum_{i\ne j}\MV u_{ij}\BAR{\HC P_j} S = m P_i.
\]
By defining 
\begin{equation}
  \label{eq-25}
  P_i^+ \stackrel {def}= P_i +\BAR{\HC P_i} \MBOX{and} 
  P_i^- \stackrel {def}= (P_i -\BAR{\HC P_i})S 
\end{equation}
 and addition/subtraction of both
equations one gets:\FN{Since $P^+$ and $P^-$ by their definition obey: 
$\BAR {\HC {(P^+)}} = P^+$ and $\BAR{\HC {(P^-)}} = - P^-$ this is a  
{\em quaternionic} decomposition of $P$ as $P = P^+ -P^- S$, where $P^+$ and $iP^-$ 
are quaternions (see appendix).
Since also $i\MV u$ is a quaternion, this set of equations (\ref{eq-26}) can represented 
with these algebraic entities.
}
\begin{equation}
 \label{eq-26}
(\varepsilon_i-m) P_i^+ = -\sum_{i\ne j}\MV u_{ij}P_j^-\MBOX{and}
(\varepsilon_i+m) P_i^- = \sum_{i\ne j}\MV u_{ij}P_j^+.
\end{equation}
The {\sc lagragian} sum (\ref{eq-19}) reads with this substitution:\FN{
  Of course, one also can derive again (\ref{eq-26}) from (\ref{eq-27}). 
} 
\begin{equation}
 \label{eq-27}
 \LAGR = \sum_i \left((\varepsilon_i-m) |P_i^+| - (\varepsilon_i+m)  |P_i^-|\right)
 + \sum_{ij}\TR{\BAR{P_i^-} \MV u_{ij} P_j^+}.
\end{equation}
At this stage the {\sc schrödinger}-approximation is easily feasible, by setting 
${\varepsilon_i+m\approx 2m}$ in  (\ref{eq-26}) or (\ref{eq-27}).
This is done in section \ref{sec-schroedinger}.
 However, in the following sections I want to present precise results.

\subsection{Ground State}
At first, I start with the simplest case: the ground state in an atom,
and will show, that the above formulas lead to the correct energy (and atom radius).\\
This state will be described with only two points ($n = 2$)\FN{
  It is easy to show, that for $n=1$ no stationary solution exists.
}: $\MV x_1, \MV x_2$. Then exists only one edge 
$\MV u_{12} = (\MV x_1- \MV x_2)^{-1} \stackrel {def}= \MV u = - \MV u_{21} $. However,
as explained in the section \ref{sec-lag-sphere}, this edge is counting {\em twice}, giving
a factor of $2$, so equation (\ref{eq-26}) reads:\FN{
  with $r_i \stackrel {def}= ||\MV x_i|| = \sqrt {-|\MV x_i|}$ and again 
  $\varepsilon_i \stackrel {def}= \varepsilon + \frac \alpha{r_i}$
}
\begin{eqnarray}
 \label{eq-28}
 (\varepsilon_1-m) P_1^+ = -2\MV u P_2^- &\MBOX{and}& (\varepsilon_1+m) P_1^- = +2\MV u P_2^+,\nonumber\\ 
 (\varepsilon_2-m) P_2^+ = +2\MV u P_1^- &\MBOX{and}& (\varepsilon_2+m) P_2^- = -2\MV u P_1^+.
\end{eqnarray}
In these four equations only the two spinor-pairs $P_1^+, P_2^-$ and  $P_2^+, P_1^-$ are coupled, 
therefore eliminating the $P_i^-$, results in the pair of equations:
\begin{equation}
 \label{eq-29}
(\varepsilon_1-m)(\varepsilon_2+m) P_1^+ = 4\MV u^2 P_1^+, \quad
(\varepsilon_1+m)(\varepsilon_2-m) P_2^+ = 4\MV u^2 P_2^+.
\end{equation}

In these two equations the spinors $P_i^+$ are then freely variable and can be divided out
(except for the two singular cases $P_1^+ = P_2^- = 0$ or $ P_2^+ = P_1^- = 0$ resp.)\FN{
  The above decomposition  on $P$ into $P^\pm$ in (\ref{eq-25}) has the consequence that the 
  \Lagrangian  (\ref{eq-27})
  becomes a sum of two terms: 
  $\LAGR_1(P_1^+,P_2^-, \varepsilon,\MV x_1, \MV x_2 ) +  
  \LAGR_2(P_2^+,P_1^-, \varepsilon,\MV x_1, \MV x_2)$,
  where the spinor-pairs must be viewed as varying independently. Therefore, these singular cases 
  not have to be considered.\\ 
  If one nevertheless computes these cases in full detail, eg. the first one,
  where only the second eq. of (\ref{eq-30}) holds, it turns out, however, that the resulting expression
  $\varepsilon(\MV x_1, \MV x_2)$ has only {\em one stationary point}, which is no extremum,
  but a saddle-point.
} and one gets\FN{
  The factors on both sides are simple reals. Consider again, that $\MV u$ is a traceless 
  matrix by definition,
  so $\BAR {\MV u} = -\MV u$ and $\MV u^2 = -|\MV u| \ge 0$.
}

\begin{equation}
  \label{eq-30} 
  (\varepsilon_1-m)(\varepsilon_2+m) = 4\MV u^2, \quad
  (\varepsilon_1+m)(\varepsilon_2-m) = 4\MV u^2.
\end{equation}
Consequently follows from $(\varepsilon_1-m)(\varepsilon_2+m) = (\varepsilon_1+m) (\varepsilon_2-m)$, that
$\varepsilon_1 = \varepsilon_2$ must hold, i.e. $r_1 = r_2 \stackrel {def}= r$.\\
The resulting equation is
\begin{equation}
  \label{eq-31} 
  (\varepsilon +\frac \alpha r)^2 -m^2 =  4\MV u^2.
\end{equation}
Since $|\MV x_1| = |\MV x_2| = -r^2$ holds, by triangle formulas one gets:
$-|\MV x_1 - \MV x_2|  = 4(r^2 - h^2)\leq 4r^2 $, with $h \leq r$ as height on the edge, so 
$4\MV u^2 = 1/(r^2 - h^2)$ and it results\FN{
  In this formula $r,h$ should not be misunderstood as usual variables: they get fixed values after using 
  the extremal principle, also $\varepsilon$ is then a constant, of course.
}
\begin{equation}
 \label{eq-32}
 \varepsilon(r,h) = - \frac \alpha r \pm \sqrt{ m^2 + \frac 1{r^2 - h^2}}.
\end{equation}
As explained above, the \Lagrangian extremal principle requires that this expression is to 
make stationary with respect to $r$ and $h$, which after simple computations
gives\FN{ 
  only the $+$ sign of the root gives for positive $\alpha$ (attractive potential)
  an extremum
} 
immediately $h \stackrel != 0$ and 
$r \stackrel != \frac {\sqrt {1-\alpha^2}}{\alpha m} \approx \frac 1 {\alpha m}$ 
({\sc bohr}s formula for the radius of the hydrogen atom) 
and finally the correct energy of the 
ground state:
\begin{equation}
 \label{eq-33}
 \underline {\varepsilon = m\sqrt {1-\alpha^2}}.
\end{equation}

\subsection{Space-Grid for the General State}
\label{sec-general-grid}
The purpose of this section is, to present a general space grid, that is {\em stationary} together
with the spinors. And I will show, that this represents the correct 
quantum states of an atom.\\ 

Therefore, I first consider a separation of variables, namely the radial variable $r$ and 
angular variables (on the sphere). This separation is possible due to the symmetry of
the {\sc coulomb} potential and is similar to usual procedures.\\

However, I want to emphasize here, that it should be possible to find 
{\em more general methods}, which do not rely on the assumption of {\em separability}, used below.
I think, that once the edge scheme is fixed (i.e. which nodes are connected by edges), it
can be proved, that the nodes are general stationary points. This was e.g. shown for
the ground state in the last section, where it was forced by the equations, that 
both nodes have the same distance to the center.\\

The grid should consist of $n$ spheres with the radii $r_i = r_1,..,r_n$, and 
all these spheres should have the same set of node normals. That means that every point of the
grid can be represented as $\MV x_k = r_i\MV p_j$, where $\MV p_j$ are unit vectors: 
$|\MV p_j| = -1$.\\
 Two points on different spheres $r_i \ne r_j$ also should only be connected by an edge, if they
have the same spherical coordinates.

Again, the task is to find a stationary point for the \Lagrangian sum in (\ref{eq-27}):
\[
  \LAGR (r_i,\MV p_i, P_i^+,P_i^-).
\]
The separation ansatz now assumes, that also the spinors $P_i^\pm$ can be factorized, namely as
\begin{equation}
  P_i^+ = f_i A_i \MBOX{and} P_i^- = g_i \MV p_i A_i,
\end{equation}
where $f_i$ and $g_i$ should be real constants only depending on the radial index and 
$A_i$ matrices, only depending on the angular index.\\

Considering that on radial edges holds $\MV x_i - \MV x_j = \MV p_i(r_i-r_j)$, introducing the
auxiliary parameters $\varepsilon_i^\pm \stackrel {def}= \varepsilon_i \pm m$ and inserting the above
into the sum (\ref{eq-27}) leads to (the summation labels $R,S$ should denote summation over
radial, resp. angular indicees):

\begin{eqnarray}  
  \LAGR &=& 
  \left(\sum_{k \in S} |A_k|\right)  \left( \sum_{i\in R}\varepsilon_i^-f_i^2 + \varepsilon_i^+g_i^2 +
  \sum_{ij \in R} \frac {f_ig_j- g_i f_j}{r_i-r_j}\right)\nonumber\\
  &+& 
   \left( \sum_{kj\in S} \TR{\bar A_k \bar\MV p_k (\MV p_k - \MV p_j)^{-1} A_j} \right)
  \left( \sum_{i\in R} \frac {f_ig_i} {r_i}\right) 
\end{eqnarray}
The usual separation idea is now, that the angular and radial dependent factors in both summands
must be separable, that requires with some constant $\kappa$:
\begin{equation}                             
  \label{eq-36}
  2\kappa \sum_S |A_i| \stackrel != \sum_S \TR{\bar A_i \bar\MV p_i (\MV p_i - \MV p_j)^{-1} A_j}.
\end{equation}
Then the sum can be decomposed into two independent factors $\LAGR = \LAGR_S \LAGR_R$, where $2\kappa$
arises as eigenvalue in $\LAGR_S$
and the radial factor $\LAGR_R$ becomes:
\begin{equation}
  \label{eq-37}
  \LAGR_R = \sum_i(\varepsilon_i^-f_i^2 + \varepsilon_i^+g_i^2)
  -2\kappa\sum_i \frac {f_ig_i} {r_i}+ \sum_{j\ne i} \frac {f_ig_j- g_i f_j}{r_i-r_j}.
\end{equation}
The solution of the angular part $\LAGR_S$ is given in the appendix, section \ref{sec-lag-sphere}.

\subsection{Solution of the Radial Equations}
By variating the $f_i,g_i$ in (\ref{eq-37}) one gets:\FN{
  Note, that the sums are only over the radial index, from here on.
} 
\begin{equation}
  \label{eq-38}
  \varepsilon_i^-f_i = \frac \kappa{r_i}g_i - \sum_{j\ne i} \frac {g_j}{r_i-r_j} \MBOX{and}
  \varepsilon_i^+g_i = \frac \kappa{r_i}f_i + \sum_{j\ne i} \frac {f_j}{r_i-r_j}.
\end{equation}

At this point, I want to emphasize the correspondence to the radial differential equations, 
derived from {\sc dirac}s-equation, with similar presumptions, namely they read:
$ \varepsilon^- f = \frac \kappa{r}g -  g', \;
 \varepsilon^+ g = \frac \kappa{r}f + f' $ (see e.g. \cite{LL}).\\
The detailled discussion of this and also the connection of the both associated 
{\sc Lagrang}ians is given in the appendix.\\

The {\em second variation}, which considers the $r_i$, additionally gives (for all $r_i$)\FN{
  Consider, that $d\varepsilon^\pm_i/dr_i = -\alpha/r_i^2$ and the double sum contains each term
  twice.
}
\begin{equation}
  \label{eq-39}
  \frac {d\LAGR}{dr_i} = -\frac \alpha{r_i^2}(f_i^2 + g_i^2)
  +2\kappa\frac {f_ig_i} {r_i^2}- 2\sum_{j\ne i} \frac {f_ig_j- g_i f_j}{(r_i-r_j)^2}\stackrel != 0.
\end{equation}
 
The equations (\ref{eq-38}) set up a system of linear equations, which can be considered as 
eigenvalue problem for $\varepsilon$ (if all $r_i$ are fixed). Together with (\ref{eq-39})
they form a set of $3n$ equations for the $3n+1$ variables $r_i,f_i,g_i,\varepsilon$.\FN{
From here on, $n$ denotes the number of spheres, not nodes.
}\\
Since the first system is linear in $f_i,g_i$ and the second bilinear, however, 
they are normalizable, consequently 
the number of equations equals the number of variables, indicating that only discrete solutions
exist.\\

For $n=1$ the solution can be derived directly, giving (correctly)\FN{
Consider from (\ref{eq-38}): $\varepsilon_1^-f_1 = \frac \kappa{r_1}g_1$, and 
$\varepsilon_1^+g_1 = \frac \kappa{r_1}f_1$ and from (\ref{eq-39})  
$ \alpha(f_1^2 + g_1^2) = 2\kappa{f_1g_1}$. This system of 3 equations has only the above solution. 
}
\begin{equation}
\label{eq-40}
\varepsilon = \frac m\kappa \sqrt {\kappa^2-\alpha^2},\quad r_1 = \frac \kappa{\alpha m}\sqrt {\kappa^2-\alpha^2}.
\end{equation}

For the general case, it turns out, that a simple linear ansatz for the $f_i,g_i$, where $a,b,c,d$ are real constants
 \begin{equation}
\label{eq-41} f_i = a + b r_i \MBOX{and} g_i = c + d r_i,
\end{equation}
and a set of $r_i$, obeying the equations (with two parameters $\lambda,\gamma$)\FN{
  These equations are
  related to {\sc laguerre}-polynoms, and discussed in the appendix.
}
\begin{equation}
\label{eq-42} \sum_{j\ne i} \frac 1{r_i-r_j} = \lambda - \frac \gamma{r_i}
\end{equation}
 gives a solution which results (after longish computations) in the correct formula 
 for the energy levels, where
$n_r = n-1$ is the radial quantum number (see e.g. \cite{LL}, p. 126):\FN{
  The parameters are then determined as
  $\gamma = \sqrt{\kappa^2 -\alpha^2}$ and $\lambda = \sqrt{m^2 -\varepsilon^2}$.
}
\begin{equation}
  \label{eq-43} 
  \Big(\frac m\varepsilon \Big)^2 = 1 + 
  \Big(\frac \alpha{\sqrt {\kappa^2-\alpha^2} + n_r} \Big)^2.
\end{equation}

\section{Electron under {\sc Lorentz}-Force}
\label{lorentz-force}
This section is intended, to demonstrate the working of the method 
for one non-stationary case.\FN{To tackle problems of this type as initial
  value problem  (e.g. numerically), one should consider the following method. 
  Suppose the three consecutive points $\MV x_1,\MV x_2, \MV x_3$ are used (and
  the associated spinors $P_1,P_2$). Then the variation equations for the inner point $\MV x_2$
  and $P_1,P_2$, (\ref{eq-45}, \ref{eq-47}) are used, building a set of 3 implicit equations. 
  In this set, the initial values of 
  $\MV x_1,\MV x_2,P_1$ are inserted, which eventually results in 
  the values for $\MV x_3,P_2$, and so forth.
}\\

I consider here the same case as in eq. (\ref{eq-7}), except that also
an electromagnetic field is present, i.e. again with $n=1$ (no space-like edges) and 
$\MV v_k \stackrel {def}= (\MV x_{k+1} -\MV x_{k})^{-1}$:\FN{
  Again this sum is obviously invariant under the "local" 
  transformations $P_k \to P_k S_k$, when $S_k\HC S_k = 1$ and
  $|S_k| = 1$. I.e. the spinors $P_k$ are determined only up to these factors by the following
  equations. This is a partial analogy to gauge invariance of standard {\sc dirac} theory, except
  the vector field $\MV A$ is not transformed here.
}
\begin{equation}
  \label{eq-44}
  \LAGR =  \sum_{k} \TR{(\MV v_k + \frac e2(\bar {\MV A}_{k+1} + \bar {\MV A}_k))P_{k}\HC P_{k} } 
  - 2 m \sum_{k} \Re(|P_{k}|).
\end{equation}
The ``field equations'' become similarly:
\begin{equation}
  \label{eq-45} 
  (\MV v_k + \frac e2(\bar {\MV A}_{k+1} + \bar {\MV A}_k))P_k = m \BAR{\HC P_k}.
\end{equation} 
By taking the determinant on both
sides of this equation immediately follows, that $|P_k|$ must be real\FN{
  consider $|\MV v + e\MV A||P| = m^2 |P|^\ast$ and $|\MV u|=real$ for any 
  {\sc hermite}an $\MV u$.
}  and 
consequently it must hold for all $k$ (it is closely related to the conservation of energy):
\begin{equation}
\label{eq-46}
|\MV v_k + \frac e2(\bar {\MV A}_{k+1} + \bar {\MV A}_k)| = m^2 = const.
\end{equation}
One remarkable consequence of this simple  formula is, that regardless of the history of a particle,
in case of vanishing vector potential, in its rest frame always holds $\Delta t = 1/m$.\FN{
  The equation, however, reveals some important new issues:\\ 
  E.g. adding a constant offset to $\MV A$ (which does not affect the classical {\sc lorentz}-force)
  here changes the discretization and thus modifies the results. It seems, that apparently no full gauge
  invariance can be derived for this model. For small fields, however, the results are equal to
  the classical theory.

Then for example, consider the simplest case:
a resting particle in scalar potential $\MV A = U(\MV x)$. Then eq. (\ref{eq-46}) reads: 
$(\frac 1{\Delta t} + U)^2 = m^2$ i.e. $\Delta t = \frac 1{|m| -U}$. Since $\Delta t > 0$ is supposed,
only the range $-\infty < U < |m|$ for the external field is possible.
}\\

The {\em second variations} (for $\MV x_k$) result in:\FN{consider again 
$\VAR \MV A_k = \frac 12\TR{\VAR \MV x_k \bar\partial}\MV A_k$
}
\begin{equation}
\label{eq-47}
\MV v_kP_k\HC P_k\MV v_k - \MV v_{k-1}P_{k-1}\HC P_{k-1}\MV v_{k-1} + 
 \frac e4 \bar\partial\TR{\bar \MV A_k(P_k\HC P_k + P_{k-1}\HC P_{k-1})} \stackrel != 0.
\end{equation}
Please, consider again, that the set of equations (\ref{eq-45}) and (\ref{eq-47}) must
be solved simultaneously.\\
By multiplying  (\ref{eq-45}) from the right with $\HC P_k$ and $(\MV v + e\MV A)^{-1}$
from left one gets\FN{if the approximation of small field  $|e \MV A| << |\MV v|$ is used.
}
\begin{eqnarray}
 \label{eq-48} 
P_k\HC P_k &=& m |P_k|(\MV v_k +\frac e2(\bar \MV A_{k+1} + \bar {\MV A}_k))^{-1}\nonumber \\
&\approx& m |P_k|(\MV v_k^{-1} - \MV v_k^{-1}\frac e2(\bar \MV A_{k+1} + \bar {\MV A}_k) \MV v_k^{-1}).
\end{eqnarray} 
Inserting this and the corresponding term for $P_{k-1}\HC P_{k-1}$ 
in (\ref{eq-47}) and omitting terms $\sim e^2\MV A^2$ gives (after division by $m$):
\begin{eqnarray}
\label{eq-49}
 |P_k|(\MV v_k  - \frac e2(\bar {\MV A}_{k+1} + \bar {\MV A}_k))
&-& |P_{k-1}|(\MV v_{k-1}  - \frac e2(\bar {\MV A}_{k} + \bar {\MV A}_{k-1}))\nonumber \\ 
+ \frac e4 \bar\partial\TR{ \bar\MV A_k (|P_k|\MV v_k^{-1} &+& |P_{k-1}|\MV v_{k-1}^{-1})} = 0. 
\end{eqnarray}
Reordering gives (after a bar-operation):
\begin{eqnarray}
\label{eq-50}
  |P_k|(\bar \MV v_k - \frac e2({\MV A}_{k+1} &+& {\MV A}_k) + 
\frac e4\partial\TR{ \bar\MV A_k\MV v_k^{-1}} )\\ =
 |P_{k-1}|(\bar \MV v_{k-1}-\frac e2({\MV A}_{k} &+& {\MV A}_{k-1}) -\frac e4\partial\TR{\bar\MV A_k\MV v_{k-1}^{-1}})
\nonumber.
\end{eqnarray}
Now with $\MV v_k^{-1} = \MV x_{k+1} - \MV x_{k}$ (by definition) the approximations are used:\FN{
  see appendix, chapter "Differential Calculus" for explanation.
}
\begin{equation}
\label{eq-51}
\MV A_{k + 1} \approx  \MV A_{k} +\frac 12 \TR{\MV v_k^{-1} \bar \partial}\MV A_k\MBOX{and}
\MV A_{k - 1} \approx  \MV A_{k} -\frac 12\TR{\MV v_{k-1}^{-1} \bar \partial}\MV A_k
\end{equation}
resulting in
\begin{eqnarray}
\label{eq-52}
  |P_k|(\bar \MV v_k - e{\MV A}_{k} &-& \frac e4 \TR{\MV v_k^{-1} \bar \partial} \MV A_k) + 
  \frac e4 \partial\TR{\bar\MV  A_k\MV v_k^{-1}} )\\ =
 |P_{k-1}|(\bar \MV v_{k-1}  - e{\MV A}_{k} &+& \frac e4\TR{\MV v_{k-1}^{-1}\bar\partial} \MV A_k 
 -\frac e4 \partial\TR{ \bar \MV A_k\MV v_{k-1}^{-1}})
\nonumber.
\end{eqnarray}
Now the equation for the fieldtensor $F$ (at point $\MV x_k$) is used 
(see appendix \ref{sec-electrodynanmics}), 
which for any $\MV u$ obeys: 
$ \MV uF + \HC F \MV u = \TR{\bar\partial\MV u}\MV A - \partial\TR{\MV A\bar \MV u}$ giving
\begin{eqnarray}
\label{eq-53}
  |P_k|(\bar \MV v_k -  e{\MV A}_{k} &-& \frac e4 (\MV v_k^{-1}F_k + \HC F_k\MV v_k^{-1} ))\\ =
 |P_{k-1}|(\bar \MV v_{k-1}  -  e{\MV A}_{k} &+& 
 \frac e4 (\MV v_{k-1}^{-1}F_{k} + \HC F_{k}\MV v_{k-1}^{-1} ))
\nonumber.
\end{eqnarray}

Since one cannot generally claim $|P_k| = |P_{k-1}|$ (which case could be easily solved), the
symmetrical ansatz (which is always possible, of course) with a new real variable $\lambda$:
$|P_k| = 1+\lambda, |P_{k-1}| = 1-\lambda $ is used. 
Also, the centered difference 
$\MV q \stackrel{def}=  \MV v_k^{-1} + \MV v_{k-1}^{-1} =\MV x_{k+1} - \MV x_{k-1}$ is used.
Then one gets:\FN{
  since $|\lambda| \ll 1$ is supposed to be small, only the dominant term 
  $\sim \lambda$ is considered, which is $\bar \MV v$
}
\begin{equation}
\label{eq-54}
  (1+\lambda) \bar \MV v_k -  (1-\lambda)\bar\MV v_{k-1}  = \frac e4 (\MV qF_k + \HC F_k \MV q ).
\end{equation}
To determine $\lambda$, this equation is multiplied (from right) with 
$\bar \MV q = \bar \MV v_k^{-1} + \bar \MV v_{k-1}^{-1}$ and then
is taken the trace, so the right-hand side vanishes (since $F + \bar F = 0$). It remains
\begin{equation}
  \label{eq-55} 
  \TR{  (1+\lambda) (1 + \bar \MV v_k\bar \MV v_{k-1}^{-1}) -  
    (1-\lambda)( 1 + \bar\MV v_{k-1}\bar \MV v_{k}^{-1}) } = 0.
\end{equation} 
Then, with $\MV a \stackrel {def}= \bar \MV v_k - \bar \MV v_{k-1}$ and another auxiliary matrix 
$B \stackrel {def}=\bar \MV v_k\bar \MV v_{k-1}^{-1} = 1 + \MV a\bar \MV v_{k-1}^{-1} $, 
one gets: 
\begin{equation}
  \label{eq-56}
   (1+\lambda)  \TR{1 + B} = (1-\lambda)\TR {1 + B^{-1} } \MBOX{i.e.}
  \lambda = \frac{\TR{B^{-1}} -\TR{B} } {4 +\TR{B^{-1}} +\TR{B}}.
\end{equation} 
With $\TR {B^{-1}} = \frac{\TR{B}}{|B|}$  and 
$|B| = \frac{|\MV v_k|}{|\MV v_{k-1}|}$\FN{
  consider from eq. (\ref{eq-46}) 
  $ |\MV v_k + \frac e2(\bar {\MV A}_{k+1} + \bar {\MV A}_k)| =
  |\MV v_{k-1} + \frac e2(\bar {\MV A}_{k} + \bar {\MV A}_{k-1})|\;\; (= m^2)$ and $|e \MV A| \ll |\MV v|$
  thus $|\MV v_k|\approx |\MV v_{k-1}|$ holds.
} 
and the approximation $\TR B \approx 2$ follows 
$\lambda \approx \frac{|\MV v_{k-1}| -|\MV v_{k}|}{4|\MV v_{k}|}$.\\

To state the approximation of {\sc lorentz}-force of eq. (\ref{eq-54}) it remains to explain, that the 
relativistic velocity vector $\MV u = d\MV {x}/d\tau = \dot \MV x$ is discretized as\FN{
  centered around $\MV x_k$,
  using $\Delta\tau = \sqrt{|\MV x_{k+1} - \MV x_{k-1}}| =
  \sqrt{|\MV v_k^{-1} + \MV v_{k-1}^{-1}|} \approx 
  \sqrt{4|\MV v_k^{-1}|} \approx 2/m$.
}
${\MV u = (\MV x_{k+1} - \MV x_{k-1})/\sqrt{|\MV x_{k+1} - \MV x_{k-1}|}} \approx \frac m2 \MV q $
i.e. $\underline{\MV q \approx \frac 2m \MV u}$ and $\MV a$ is actually a discretized acceleration vector:
\begin{eqnarray}
  \label{eq-57} 
  \MV a =\bar \MV v_k - \bar\MV v_{k-1} 
  &=& |\MV v_k| (\MV x_{k+1} - \MV x_k) -|\MV v_{k-1}| (\MV x_{k} - \MV x_{k-1}) \nonumber \\
  &\approx & \frac  {\MV x_{k+1} - 2\MV x_k +\MV x_{k-1}} {(\Delta\tau)^2} 
  \approx \frac {d^2 \MV x}{d\tau^2}. 
\end{eqnarray}
Finally from eq. (\ref{eq-54}) results the equation of {\sc lorentz}-force (see appendix) with 
small corrections:\FN{
  The correction term resembles a corresponding term in {\sc dirac}s motion equation, which reads
  in this notation and scaling: $\frac{e^2}{6\pi m} (\dot \MV a - \MV u |\MV a|)$ 
  (see e.g. \cite{HS-SRT}, p. 173)
}
\begin{equation}
  \label{eq-58}
  \underline{ \MV a = \frac{e}{2m}(\MV u F_k + \HC F_k\MV u)} - 2\frac \lambda m \MV u.
\end{equation}

\section{{\sc schrödinger}-Approximation and ``{\sc hamilton}-Sum''}
\label{sec-schroedinger}
This section shall demonstrate, that also a {\em discrete form} of the classical stationary 
{\sc schrödinger} equation and its associated {\sc hamilton}ian can be derived as 
approximation from the above discrete {\sc dirac} formalism. This approximation
is always possible for electrons in weak electromagnetic fields.\FN{
  The way of deriving this approximation from {\sc Dirac} equation is similar to standard QM.
}\\

I start with equation (\ref{eq-26}) from section \ref{sec-coulomb}, 
which represents a bound state of an electron, 
but here in a general electric potential field $V(\MV x)$ (real scalar, time independent) 
with  $V_i \stackrel {def} = V(\MV x_i)$:\FN{and again
$ \MV u_{ij} \stackrel {def}= (\MV x_i - \MV x_j)^{-1}$ for the space-edges
}
\begin{equation}
 \label{eq-59}
(\varepsilon + V_i - m) P_i^+ = -\sum_{j}\MV u_{ij}P_j^-\MBOX{and}
(\varepsilon + V_i +m) P_i^- = \sum_{j}\MV u_{ij}P_j^+.
\end{equation}
As mentioned, herein  the {\sc schrödinger}-approximation is easily feasible, by setting 
${\varepsilon + V_i +m\approx 2m}$ in the second equation.\FN{This is the usual approximation method
for small energy, since $\varepsilon \approx m$ and $V \ll m$}
Then $P_i^-$ can be expressed directly with it:
\begin{equation}
  \label{eq-60}
  P_i^- \approx \frac 1{2m} \sum_{j}\MV u_{ij}P_j^+.
\end{equation}
Inserting this in the first of (\ref{eq-59}), gives with $E \stackrel {def} = \varepsilon -m $ as energy:\FN{
  in many textbooks the potential energy $U = -V$ is used instead of $V$ in the formulas
}
\begin{equation}
  \label{eq-61}
  (E + V_i) P_i^+ =  -\frac 1{2m} \sum_{jk}\MV u_{ij}\MV u_{jk} P_k^+.
\end{equation}
Please note, that the double-sum is to build over all edge-pairs $(i,j)$ and $(j,k)$. This is still
a {\em matrix equation}. To get a {\em scalar} equation from it, one adds the adjuncted, resulting 
in a scalar expression:\FN{consider that $E,V_i$ are scalars and
  $\bar \MV u_{jk} = -\MV u_{jk}$ and $\MV u_{kj} = -\MV u_{jk}$ holds.
}
\[ 
  (E + V_i) (P_i^+ + \bar {P}_i^+)  = 
  -\frac 1{2m} \sum_{jk} (\MV u_{ij}\MV u_{jk} P_k^+ + \bar P_k^+\MV u_{kj}\MV u_{ji} ),
\] 
and then drops the vector part of $P_k^+$ in the r.h.s. (ignoring all spin-effects), by setting it
a real scalar $\psi_k \stackrel{def}= P_k^+ = \bar P_k^+ $.\\
The result is a discretized form of the stationary {\sc schrödinger}-equation:
\begin{equation}
  \label{eq-62}
  (E +V_i) \psi_i =  -\frac 1{2m} \sum_{jk}\frac 12(\MV u_{ij}\MV u_{jk} + \MV u_{kj}\MV u_{ji}) \psi_k.
\end{equation}
The sum $\frac 12 \sum_{jk}(\MV u_{ij}\MV u_{jk} + \MV u_{kj}\MV u_{ji})\psi_k $ there 
represents the second order partial derivation operator $\Delta\psi_k$\FN{
  It is always scalar and real, of course.
} at the point $\MV x_i$.\\

One can also easily define a "{\sc hamilton}ian-sum", from which the 
above {\sc schrödinger}-equation (\ref{eq-62})
can be derived again (by variation of all $\psi_k)$:\FN{
  Of course, it it also possible to derive this {\sc hamilton}ian directly from
  the \Lagrangian sum, with the same assumptions.
}
\begin{equation}
  \label{eq-63}
  H(\psi_i, \MV x_i) = \sum_i(E +V_i) \psi_i^2 + 
  \frac 1{2m} \sum_{ijk}\MV u_{ij}\MV u_{jk} \psi_k\psi_i 
\end{equation}
This sum obviously corresponds to the classical {\sc hamilton}ian for the stationary case.\FN{
  The proof, that the triple sum is always a real scalar, is simple when again 
  the hermitecity and antisymmetry of the $\MV u_{ij} = \HC{\MV u_{ij}} = - \MV u_{ji}$ is used,
  e.g. $\HC{(\MV u_{12} \MV u_{23})} = \MV u_{23}\MV u_{12} = \MV u_{32}\MV u_{21}$.
}
\subsection{General Considerations about Ground States}
At first, I want to discuss the solution for the ground states of any potential $V$, 
which is described here with $ n = 2$ nodes.\\
Then one has only one edge with
$ \MV u \stackrel {def}= \MV u_{12} =  (\MV x_1 - \MV x_2)^{-1} = -\MV u_{21}$ and two equations:
\begin{equation}
  \label{eq-64}
  (E + V_1)\psi_1 =  - \frac 1{2m} \MV u_{12} \MV u_{21} \psi_1  = \frac {\MV u^2}{2m} \psi_1 
  \MBOX{and} (E + V_2)\psi_2 = \frac {\MV u^2}{2m} \psi_2.
\end{equation}
It follows $  E + V_1 = E + V_2 = \frac {\MV u^2}{2m} $, i.e. at first one can
conclude $V_1 \stackrel != V_2$.\\
Please note, that for every {\sc minkowski}an pure space-vector $\MV u$ 
(with ${\bar \MV u = -\MV u}$) holds $\MV u^2 = - |\MV u| > 0$, therefore $E > -V_{1,2}$\\
Also notable is the decoupling of $\psi_1,\psi_2$ in equation (\ref{eq-64}), meaning
that their  values are independent.\FN{
  This does not hold for the exact solution, given in chapter 4.2., there the spinor $P_{k}$
  cannot vanish at any point $k$.
}\\

However, in contrast to the case of  {\sc Coulomb}-potential discussed below: 
if there exists a stationary point $\MV x_0$ of the potential 
(with $\frac{\partial V(\MV x_0)}{\partial \MV x} = 0$), then also
a solution with only one node ($n=1$, no edge) is possible, which then would have the energy
$E = - V(\MV x_0)$.\\
These solutions do not have counterparts in current QM. To exclude them, there must
be a principle, that excludes stationary solutions with $n=1$. At the moment I cannot see,
what this can be. 

\subsection{Ground State in {\sc Coulomb}-potential}
The {\sc Coulomb}-field (of an atom nucleus) is (with $r = ||\MV x||$ as euclidian distance) 
\begin{equation}
  \label{eq-65}
  V(\MV x) = \frac \alpha r.
\end{equation}
Then (again from $V_1 = V_2$) directly follows $ r_1 = r_2 = r$. 
As explained above, in contrast to the one-dimensional case, the edge is counting {\em twice} again,
so one gets
\begin{equation}
  \label{eq-66}
  E = - \frac  \alpha {r} +\frac {4\MV u^2}{2m}
\end{equation}
By setting $1/ \MV u^2 = (\MV x_1 - \MV x_2)^{2} = 4(r^2-h^2)$ results:\FN{
  With simple triangle formula for the triangle ($\MV x_1, \MV 0, \MV x_2$), with $h$ as height
  and $h \le r$.
} 
\begin{equation}
  \label{eq-67}
  E(r,h) = -\frac  \alpha {r} +\frac 1{2m(r^2-h^2)}.
\end{equation}
The condition $ \PD [E] h = 0$ then gives $h = 0$, and $\PD [E] r = 0$ gives
$r = 1/\alpha m$ (atom radius) and finally the energy of the ground state of hydrogen:
\begin{equation}
  \label{eq-68}
  E = - m \frac {\alpha^2}{2}.
\end{equation}

\subsection{Quantum Harmonic Oscillator}
This section is included, to give readers the most simple testcase of the discrete theory.
The one-dimensional harmonic oscillator has the field (with usual scale factor):
\begin{equation}
  \label{eq-69}
  V(x) = - \frac m2 \omega^2x^2.
\end{equation}
Here for simplification  is set $m = \omega = 1$, giving from (\ref{eq-63}) 
the ``{\sc Hamilton}ian sum'':\FN{
  Consider e.g. the one-dimensional case with $\MV x = {0, x\choose x,0}$ ($y=z=0$).
  Then the $\MV u_{ij} = ({\MV x_i - \MV x_j})^{-1}$  
  all commute, and their products $\MV u_{ij}\MV u_{jk} = \frac {1}{(x_i-x_j)(x_j - x_k)} $ 
  are always simple scalars
}
\begin{equation}
  \label{eq-70}
  H (\psi_i, x_i) = 
  \sum_i(E -  \frac 12 x_i^2)\psi_i^2 + \frac 12\sum_{ijk} \frac {\psi_i\psi_k}{(x_i-x_j)(x_j - x_k)}.
\end{equation}
Here all node-pairs $(i,j)$ shall be connected by one edge, with a total of ${\frac{n(n-1)} 2}$ edges.\\
The ``field equation'' (by variation of $\psi_m$) results in $n$ eqn.
\begin{equation}
  \label{eq-71}
  2(E - \frac 12 x_m^2)\psi_m + \sum_{jk} \frac {\psi_k}{(x_m-x_j)(x_j - x_k)} \stackrel != 0.
\end{equation}
The variation (differentiation) of (\ref{eq-15}) by $x_m, m = 1,..,n$ gives also $n$ eqn:

\begin{eqnarray} 
  -x_m\psi_m^2 &-& \frac 12 \sum_{jk} \frac {\psi_m\psi_k}{(x_m-x_j)^2(x_j - x_k)}
  + \frac 12 \sum_{ik} \frac {\psi_i\psi_k}{(x_i-x_m)^2(x_m - x_k)} \nonumber\\
  &-& \frac 12 \sum_{ik} \frac {\psi_i\psi_k}{(x_i-x_m)(x_m - x_k)^2}
  +\frac 12 \sum_{ij} \frac {\psi_i\psi_m}{(x_i-x_j)(x_j - x_m)^2} \stackrel != 0.\nonumber
\end{eqnarray}
This can be simplified to:\FN{ 
  by the reassignement 
  of the sum indicees in the 4. sum $i\to k$ (equals $-$ 1. sum) and of the 3.rd sum 
  $i \leftrightarrow k$ (equals $-$ 2.) and then $i\to j$
}
\begin{equation} 
  \label{eq-72}
  -x_m\psi_m^2 -  \sum_{jk} \frac {\psi_m \psi_k}{(x_m-x_j)^2(x_j - x_k)}
  + \sum_{jk} \frac {\psi_j \psi_k}{(x_j-x_m)^2(x_m - x_k)} \stackrel != 0
\end{equation}

These $2n$ equations (\ref{eq-71}, \ref{eq-72}) are to solve with $2n$ variables ($\psi_i,x_i$). 
Since it is wellknown, that {\sc Hermite}-polynoms are the eigenfunctions of the 
classical quantum harmonic oscillator, it is suggested using them to find a solution.
And indeed, it is simple to prove the following solution with their help.\FN{
  Possibly this is not the only solution, also if permutations of the $x_i$ are 
  considered.
}\\

I will show in the following, that 
with $x_i$ as zeros of these  {\sc Hermite}-polynoms, and the most simple ansatz for the $\psi_i$:
$\underline{\psi_i = a}$ (an arbitrary constant) all $2n$ equations are fulfilled, 
so this is actually {\em a stationary point} of $H(\psi_k,x_k)$.\\

The zeros of {\sc Hermite}-polynoms $(x_1,\dots, x_n)$ obey the implicit equation set\FN{
  Please note, that the $x_i$ are uniquely determined by (\ref{eq-73}) 
  (up to permutations, of course).\\
  The first few zeros can be computed explicitely, e.g. 
  for $\underline {n=1}$: $ x_1 = 0$, for $\underline {n=2}$: $x_{1,2} = \pm \sqrt{\frac 12}$, for
  $\underline {n=3}$: $x_1 = -\sqrt {\frac 32}, x_2 = 0, x_3 = \sqrt {\frac 32}$.
}
\begin{equation}
  \label{eq-73}
  \sum_{i=1, i\ne k}^n \frac 1{x_k - x_i} = x_k
\end{equation}
This is easy to prove with the methods given in appendix E: ``Orthogonal Polynoms''.\FN{ 
  They have the generating differential equation (see eq. (\ref{eq-a30})):
  $ y'' - 2xy' + \lambda y = 0 $, i.e. $u(x) = 1$, $ v(x) = -2x$, $\frac v u = -2x$.
}\\
The proof of (\ref{eq-72}) is then simple (factors $a^2$ dropped) with reordering both double sums:
\[ \sum_{j} \frac 1{(x_m-x_j)^2}\left( \sum_k\frac 1{x_m-x_k}-\sum_k\frac 1{x_j-x_k}\right) =  
  \sum_{j} \frac {x_m-x_j}{(x_m-x_j)^2} = \underline {x_m}.
\]
From the above implicit sum formulas (\ref{eq-73}) for the $x_i$, one can easily derive\FN{
  e.g. with $\sum_i \frac {x_i}{x_k - x_i} = 
  \sum_i \frac {x_i- x_k + x_k}{x_k - x_i} = \sum_{i\ne k} (-1 + \frac {x_k}{x_k - x_i}) = 
  -(n-1) + x_k^2$.
}
\[
  \sum_{i=1}^n x_i = 0\MBOX{and} \sum_{i\ne k}^n \frac {x_i}{x_k - x_i} = x_k^2 -(n-1).
\] 

The double sum in (\ref{eq-71}) then becomes:
\[ \sum_{jk} \frac {a}{(x_m-x_j)(x_j - x_k)} = \sum_j\frac {ax_j}{(x_m-x_j)} = a(x_m^2 - (n-1))
\]
and one gets:
\begin{equation}
  \label{eq-74}
  \underline {E = \frac {n-1}2}
\end{equation}
This formula reproduces the wellknown energy levels of the quantum harmonic oscillator, if the
restriction $n = even$ is made.\FN{
  Consider the units $\hbar = \omega = 1$
}\\
However, it is not clear yet, which physical principle excludes the other solutions (for $n = odd$).

\section{Conclusions and Outlook}
Here I presented a new discrete view to quantum mechanics, where the continuous wave functions
are replaced by a space-time graph with attached constant spinors and the differential equations
by discrete, algebraic equations. These equations are derived from a general ``\Lagrangian sum''
over the graph.\\

The remarkable new idea is, that the {\em graph nodes} are not to be arbitrary set,
but determined by the variation principle for {\em the same sum}.\\
Since the graph is a {\sc minkowski} space-time graph, this includes the {\em time steps} 
(which are for stationary cases then determined by the energy eigenvalue of the state) 
and gives also a valid description of {\em particle movement} (nonstationary case).\\

With this model many  classical problems of quantum mechanics are solved and give 
the expected results. 
(However, some of the solutions do not have a correspondence in classical QM. 
It is not yet clear, which {\em physical principle} suppresses them.)\\
It is thus my hope, that this model can make the wave function obsolete
(similar to the {\em light aether}, that became obsolete by the theory of Special Relativity),
and {\em all quantum phenomena can be described by a finite number of numbers, as an 
algebraic theory}, like {\sc A. Einstein} suggested.\\

However, there remain also many unanswered questions:\\
One principal task is, to introduce real dynamic
behaviour into the theory, e.g. to describe emission and absorption processes. 
Then also the model of a photon should arise.
The space-time graph for a photon, as massless particle
must be described differently, however, since the time-like edges must be replaced by 
{\em light-like} edges, with $|\Delta \MV x| = 0$.

These processes could be probably modelled  with  bifurcations and combinations of the graph.
In general, due to the implicit character of the formulas, this should be possible, because
they may have more than one solution.\\

Additionally, the usage of the electromagnetic vector potential in the theory,
can only be seen as an approximation of interactions with (virtual) photons.
This however, would imply significant changes of the variational principle.\FN{
  However, it should be stressed, that for the most important case ({\sc coulomb}-potential), there 
  exists a striking correnspondence between
  the factors of the kinematic terms in sum (\ref{eq-4}), i.e. $(\MV x_i - \MV x_j)^{-1}$ and the
  potential term $e\MV A_i = \alpha r_i ^{-1}$ (with $ r_i = ||\MV x_i - \MV x_0||$), 
  that suggests, both terms may have the same basic cause. 
} 
\\

Another interesting aspect is the question, if
it is possible, to set one primary entity (between spinors and {\sc minkowski}an vectors), from which
the other can be constructed as derivation. Apparently, this can be only the spinor part.\\
However, it is to expect that this question can only be solved in a more general framework, which
I suppose  to be a discrete theory at the {\sc plank}-scale level ({\em quantum gravity}), 
that will have a quite different concept of space-time.
From that, the presented theory will arise as approximation.\\ 

Other important tasks are:
\begin{itemize}
\item Is it possible, to simplify the \Lagrangian sum (\ref {eq-5}), e.g.
  combine time-like and space-like terms, and find some 
  kind of deeper explanation for it? 
\item Can gauge invariance represented better?
  What happens for strong fields, where $||e\MV A|| \sim m $? 
\item How are many-particle systems described? This should be possible, of course, by defining
  a composed  \Lagrangian. 
\item How are antiparticles described in this theory?
\item Is it possible, to embed this theory into the framework of general relativity? 
\end{itemize}

At last, it is to ask, of course, how far the theory is consistent with the current experimental
knowledge. Especially, the representation of particle waves ({\sc de Broglie}-waves) 
and entangled quantum states would be a challenge.

\begin{appendix}
\setcounter{equation}{0}

\section{{\sc Dirac}-Equation in Matrix-Notation versus usual Spinor-Notation}
As stated above, I will show here, that both notations are equivalent.\\
For this purpose, I start with the conventional representation for 4-spinors:\FN{
see e.g. \cite{EBERT}, pp. 24
} 
\begin{equation}
 \label{eq-a1} i\gamma_\mu \partial^\mu\psi = m\psi.
\end{equation}
There $\psi$ is a 4-column vector, and the $\gamma_\mu$ are $4\times 4$ matrices.
Here, I use the {\sc weyl}-representation for the $\gamma_\mu$:\FN{
  The 3 {\sc pauli}-matrices are again
  $\sigma_1 = {0,1\choose 1,0},\; \sigma_2 = {0,-i\choose i,\;\;0}, \;\sigma_3 = {1,\;\; 0\choose 0,-1}$
}
\begin{equation}
 \label{eq-a2} \gamma_0 = {0, -I_2\choose -I_2,0},\MBOX{and}
\gamma_k = {\;\;\; 0,\;\; \sigma_k\choose -\sigma_k, 0},\quad k=1,2,3.
\end{equation}

Then the 4-spinor is decomposable into two 2-spinors $\psi = {\Psi \choose \Phi}$, which
transform independently, but different (see below) under {\sc lorentz}-transformations,
and (\ref{eq-a1}) decomposes into a coupled system:
\begin{equation}
 \label{eq-a3} 
i(-\partial^0 +\sigma_k\partial^k)\Phi = m\Psi\MBOX{and}
i(-\partial^0 -\sigma_k\partial^k)\Psi = m\Phi.
\end{equation}
Now I define a  ({\sc hermite}an) differential operator (a $2\times 2$ matrix)\FN{
  note, that $\BAR I_2 = I_2$ and $\bar \sigma_k = -\sigma_k$ holds, 
  i.e. $\bar \partial = \partial^0 +\sigma_k\partial^k$
  and all $I_2,\sigma_k $ are {\sc hermite}an matrices
}
\begin{equation}
  \label{eq-a4} 
  \partial \stackrel{def} = \partial^0 -\sigma_k\partial^k
\end{equation}
explicitly: 
\begin{equation}
  \label{eq-a5}
  \partial = {\PD t - \PD z, \PD x +i \PD y\choose \PD x -i \PD y,\PD t + \PD z } =
  \PD t + \nabla.
\end{equation}
Then the equations (\ref{eq-a3}) read:
\begin{equation}
  \label{eq-a6}
  -i\partial \Phi = m \Psi\MBOX{and} -i\bar\partial \Psi = m \Phi.
\end{equation}
The {\sc lorentz}-transformation $T$, $|T|=1$ here operates as follows on the entities:
\begin{equation}
 \label{eq-a7}
\partial \to T\partial \HC T, \quad \bar \partial \to \bar {\HC T} \bar \partial \bar T,
\quad \Phi \to \bar {\HC T} \Phi,\quad \Psi \to T \Psi.
\end{equation}
Then both equations (\ref{eq-a6}) are obviously covariant under this transformation.\\
As usual, an electromagnetic interaction is introduced by the substitution
${ \partial^\mu \to \partial^\mu -ieA^\mu }$, which gives here:
\begin{equation}
 \label{eq-a8}
(i\partial + e\MV A) \Phi = - m \Psi\MBOX{and} (i\bar \partial +e\bar{\MV A}) \Psi = -m \Phi.
\end{equation}
Then the second equation of  (\ref{eq-a8}) is converted in the following manner.\\
One states the general formula for every 2x2 matrix $M$ ($M^T$ denoting transposed matrix): 
$\bar M = \sigma M^T \bar \sigma$, with 
$\sigma \stackrel {def}=i\sigma_2={\;\;0,\;1\choose -1,0}$.\FN{
The reason is, that the bar-operation means
spatial inversion $(x,y,z) \to (-x,-y,-z)$, and that is equal to the combined operation
of transposing (i.e. $y \to -y$) and a rotation around $y$ of 180°, given by $T = i\sigma_2$.  
}
Since  (\ref{eq-a8}b)  has the form $\bar M\Psi = -m\Phi$, with $M = i\partial + e\MV A$, 
one gets
$\sigma M^T\bar \sigma \Psi = -m\Phi$, which can be rewritten to
\FN{
  trivially, since $\bar \sigma = -\sigma, \sigma^2 = -1$
} $ M^T\sigma \Psi = -m \sigma \Phi$.

Of this one takes the complex conjugate, where 
$(M^T)^\ast = \HC M$:
\begin{equation}
 \label{eq-a9} \HC M \sigma \Psi^\ast = -m\sigma \Phi^\ast, \MBOX{with}
\HC M = -i\partial + e\MV A.
\end{equation}
One then defines a new operator for 2-spinors 
$\underline{\HCB \Psi \stackrel {def}= \sigma \Psi^\ast}$, which obeys 
$\HCB {\HCB \Psi} = -\Psi$ (since $\sigma^2 = -1$) and with that eq. (\ref{eq-a9}) then writes
$ (-i\partial + e\MV A)\HCB \Psi = -m\HCB \Phi$.\\ 
One then can combine both equations into one 2x2 matrix equation
\begin{equation}
 \label{eq-a10} 
e\MV A(\Phi,\HCB\Psi) + i\partial(\Phi,-\HCB\Psi) = -m(\Psi,\HCB\Phi).
\end{equation}
Now one defines the ``spinor-matrix'' $ P \stackrel {def}= (\Phi,\HCB\Psi)$ and 
states $\BAR{\HC P} = -(\Psi,\HCB\Phi)$, \FN{An explicit proof is
$\Phi = {\phi_1\choose \phi_2}$, $\Psi = {\psi_1\choose \psi_2}$, 
$\HCB\Psi = {\CC\psi_2\choose -\CC\psi_1}$,
 $P = {\phi_1,\;\;\CC\psi_2\choose \phi_2,-\CC\psi_1}$, 
$\HC P = {\CC\phi_1,\;\;\CC\phi_2\choose\psi_2,-\psi_1}$,
$\BAR{\HC P} = {-\psi_1,-\CC\phi_2\choose-\psi_2,\;\;\CC\phi_1} = -(\psi,\HCB\phi)$.
} and with the auxiliary matrix 
$S \stackrel {def}={i,\;\;0\choose 0,-i} $ finally gets:\FN{
  By using $S^2 = -1$ follows (e.g. by {\sc taylor} expansion):
  $ e^{\lambda S} = \cos \lambda + S \sin\lambda = {e^{i\lambda},\;\; 0 \choose 0, e^{-i\lambda }}$.
}
\begin{equation}
 \label{eq-a11}
\underline{ e\MV A P + \partial PS = m\BAR{\HC P}}.
\end{equation}
Note, that according to above definitions $P$ transforms consistently with 
${P\to \HC{\BAR T} P}$ under {\sc lorentz}-transformations and the equation ($\ref{eq-a11}$) 
is obviously covariant.\\

{\em Gauge covariance} is in this notation represented with the local transformation 
$P \to P U$, with 
$ U \stackrel {def}= e^{\lambda S}$ (where
($\lambda (\MV x)$ is  a real, scalar function of space-time). 
Then holds 
(the parentheses are set here, to denote the action of the differential operator $\partial$):
$\partial (PU) = (\partial P) U + (\partial \lambda) PUS$. Then, equation
(\ref{eq-a11}) is covariant, 
if the simultaneous transformation $e\MV A \to e\MV A + \partial \lambda$ is used.\\

The {\em stationary case} is given with the ansatz 
$P(\MV r, t) = P_0(\MV r) U(t)$, with
$ { U(t) \stackrel {def}= e^{-\varepsilon t S} }$, 
which (since $\partial_t P = -\varepsilon P S$)\FN{ 
  Since the scalar operator $\partial_t$ commutes with $P_0$ and $\partial_t U = -\varepsilon US$. 
  Also obviously $U$ commutes with $S$ and $\BAR{\HC U} = U$.
}
results in:
\begin{equation}
 \label{eq-a12}
(e\MV A + \varepsilon) P_0 + \nabla P_0S = m\BAR{\HC P_0}.
\end{equation}

\section{Relativistic Electrodynamics in Matrix-Notation}
\label{sec-electrodynanmics}
I will  shortly sketch here the basic equations of relativistic electrodynamics 
in matrix-notation without explicit proofs. Each equation can be checked, e.g. by converting
it to usual component notation.\\

The tensor of the electromagnetic field is defined from the vector potential by:\FN{
  Contrary to usual notations, here differential operators like $\partial$ can operate to the right, 
  resp. left.
  In ambiguous cases, therefore the operand  should be marked.
}
\begin{equation}
 \label{eq-a13} F \stackrel {def}= \frac 12(\bar \partial \MV A - \bar{\MV A} \partial).
\end{equation}
It is a traceless matrix ($F + \BAR F = 0$, by definition) and obeys the transformation rule 
${F \to \HC{\bar T} F \HC T}$. It can be decomposed into a {\sc hermite}an 
and anti-{\sc hermite}an part, that are the electrical and magnetical field vectors,
which both are  {\sc hermite}an, traceless matrices ($\HC E = E,\quad \HC B = B$):
\begin{equation}
 \label{eq-a14} F = E +i B\MBOX{and} \HC F = E -iB.
\end{equation}
Therefore, it is obvious, that both transform independently under spatial rotations, but
are mixed under special {\sc lorentz}-transformations.\\

The {\sc maxwell}-equations are simply (with $\MV J$ as current, also {\sc hermite}an)\FN{
these are actually 8 real equations, for the real and imaginary parts!}
\begin{equation}
 \label{eq-a15} \MV J = \partial F,
\end{equation}
and the equation of continuity (follows from last eq. with $F +\bar F = 0$) reads as 
\begin{equation}
 \label{eq-a16} \TR {\partial \BAR {\MV J}} = \partial \BAR {\MV J} + \MV J \BAR\partial = 0.
\end{equation}
Finally, the {\sc lorentz}-force on a particle with mass $m$ and electrical charge $e$, that is
moving with the relativistic velocity vector $\MV u = d\MV x/d\tau$\FN{
$\tau$ is the eigentime, given from $d\tau = \sqrt{|d \MV x|}$
} results in an acceleration vector $\MV a = d\MV u/d\tau$:\FN{The
  orthogonality of $\MV a,\MV u$ is written as 
$\TR{\MV a \bar \MV u} = 0$ and follows directly from (\ref{eq-a17}).
}
\begin{equation}
 \label{eq-a17}\underline{ \MV a = \frac e{2m}(\MV uF + \HC F \MV u)}.
\end{equation}

At last, I have to derive an identity for the last term of above equation, 
valid for arbitrary $\MV u$, which is used in the section \ref{lorentz-force}:\FN{Note, 
that $ \TR{\bar \partial \MV A} =\TR{\partial\bar{\MV A}} $.}
\begin{eqnarray}
 \label{eq-a18} 
 \MV uF + \HC F \MV u &=&  \frac 12( \MV u(\bar \partial \MV A - \bar{\MV A} \partial)
+ ( \MV A \bar \partial-  \partial\bar{\MV A})\MV u)\nonumber \\ 
&=& \frac 12(\MV u(\TR{\bar \partial \MV A} - 2\bar{\MV A} \partial)+ 
( 2\MV A\bar\partial-\TR{\partial\bar{\MV A}})\MV u) )\nonumber  \\
&=& \MV A\bar\partial\MV u -\MV u \bar{\MV A} \partial = 
 \MV A\bar\partial\MV u + \underbrace{\MV A\bar \MV u \partial - \MV A\bar \MV u \partial}_{=0}
 -\MV u \bar{\MV A} \partial \nonumber\\
&=& \TR{\bar\partial\MV u}\MV A - \partial\TR{\MV A\bar \MV u}.
\end{eqnarray}

\section{Differential Calculus and Approximations for Matrices }
The formulas stated here are standard vector analysis, they are shortly listed
here for readers, not so familiar with the notations in this paper.\\

The total differential for any field (matrix or scalar) $U$ is given simply by
\begin{equation}
 \label{eq-a19} dU(\MV x + \MV{dx}) = U(\MV x + \MV {dx}) - U(\MV x) =
\frac 12\TR{\MV {dx}\bar \partial} U.
\end{equation}
The simple explanation is, that the scalar invariant is explicitly written:
$\frac 12\TR{\MV  {dx}\bar\partial} = dt \partial_t + dx \partial_x+\cdots $\\

One often above used approximation is for the expression $(X+\delta)^{-1}$, where 
$X,\delta$ are both matrices, with $ |X| >> |\delta| $. However, the following approximation
holds for any algebra. One states\FN{
The expansion $(1-\delta)^{-1} = 1 +\delta +\delta^2 + \cdots$ can be easily checked
by multiplying both sides with $(1-\delta)$. It converges, if $\lim_{n\to \infty} \delta^n = 0$, which
is guaranteed by $|\delta| < 1$
}
\begin{equation}
 \label{eq-a20}
(X+\delta)^{-1} = (X( 1+ X^{-1}\delta))^{-1} = ( 1+ X^{-1}\delta))^{-1} X^{-1}
\approx  ( 1- X^{-1}\delta)) X^{-1}
\end{equation}

\section{Matrix-Notation and Quaternions}
Quaternions offer an elegant method for many computations, especially on the unit sphere
and generally with space rotations.\\
They are representable by the sub-algebra of matrices, obeying $\HC Q = \BAR Q$.\FN{
It is trivial, that any product of two quaternions is a quaternion again. 
}\\
The general form is obviously, with arbitrary complex $\alpha, \beta$:
\begin{equation}
  \label{eq-a21}
  Q = {\alpha, \;\; \beta \choose -\CC\beta,\CC \alpha}.
\end{equation}

To represent a pure space vector $\MV v$ with quaternions, one uses $Q = i\MV v$, which is
obviously a quaternion, since $\BAR {\MV v} = -\MV v $.\FN{
{\sc minkowski} matrices with time components, however, cannot be represented directly.
}\\
The quaternionic units $(\MV{ i,j,k})$ are consequently equal to
$(i\sigma_1,i\sigma_2,i\sigma_3)$, giving (an arbitrary real $s$ can be added, 
since it is not changed by rotations):
\begin{equation}
 \label{eq-a22}
Q = {s + iz, ix + y \choose ix-y, s-iz} = s + x\MV i + y \MV j + z\MV k.
\end{equation}
The norm (here as matrix determinant) of a quaternion is always a positive real: 
$|Q| = |\alpha|^2 + |\beta|^2 = { s^2 + x^2 + y^2 + z^2 }$.\\

Ordinary space rotations are directly represented by normalized quaternions, 
following from $\HC T = \BAR T$ (i.e. also $T\HC T = \HC T T = 1$), and 
consequently the whole apparatus of the 3-dimensional
vector space can be drawn with quaternions.

E.g. a rotation around any (space) axis $A = \mu_1 \MV i + \mu_2 \MV j +\mu_3 \MV k$ 
(with $|A| = \sum \mu_i ^2 = 1$)  by an angle $\lambda$, is represented by the transformation matrix 
$T = e^{\lambda A} = \cos \lambda + A \sin \lambda$, which is obviously a normalized quaternion. 

\section{Orthogonal Polynoms and {\sc Lagrange}-Formalism}
\label{sec-orth-poly}
This chapter is intended to illuminate the general correspondencies between eigenvalue
problems (represented by {\sc Lagrang}ians) and orthogonal polynoms.\FN{
  Orthogonal polynoms are mainly used in numerical mathematics, e.g. to compute integrals.
  Their usefulness in problem solving is widely unknown, however.
}\\
In fact, this relationship was the motivation to engage in the above theory. It shows,
that many problems of mathematical physics, described by eigenvalue problems,
can be reduced to small sets of equations for the roots of corresponding 
orthogonal polynoms.

The pair of {\sc dirac} radial equations for the 1-electron atom are used here, as especially
related example, yet there exist many other applications.\\
All following refers to the one-dimensional case, however.\FN{Thus all variables  
in this section are simple reals.}
For more than one dimension,
there will probably exist similar methods.\\ 

In the following, I will sketch some major relations for OP. Many of them 
(but not all), can be also found in standard textbooks, but derived with different formalisms.\\

\subsection{Basic Formulas for OP}
{\bf Definition:} an OP of degree $n$ : $P^{(n)}(x)$ is the (unique) polynom, associated 
with an interval $[a,b]$ and a weight function $w(x) \ge 0, x\in [a,b]$, that is orthogonal to
{\em all polynoms} $Q(x)$ of degree $q < n$, i.e.:\FN{
Thus, it roots can be computed e.g. with the following $n$ equations:
$ \int dx w P^{(n)} =\int dx w P^{(n)}x =  \int dx w P^{(n)}x^2 = \cdots = 
\int dx w P^{(n)}x^{n-1} =  0.
$}
\begin{equation}
 \label{eq-a23}
\int_a^b dx w(x) P^{(n)}(x) Q(x) = 0.
\end{equation}
These polynoms are unique up to a constant factor, of course. The roots, however, are unique. 
In the following we only deal with polynoms of the form $(x-x_1)\cdots (x-x_n)$, i.e. 
the highest coefficient is unity.
These are called {\em monic polynoms}. With this condition they are unique.\\

It follows immediately, that two OP of different degrees $n,m$:  $P^{(n)},P^{(m)} $, are orthogonal
(one is a polynom of lower degree).\\

In the following, the superscript of $P^{(n)}$ and the integral boundaries $[a,b]$ are omitted, however,
since they are considered fixed.\\

Let $P(x) = \prod\limits_{i=1}^n (x-x_i) = (x-x_1)\cdots (x-x_n)$, where all $x_i$ are real and distinct.\\

Now one defines $n$ associated {\em partial polynoms} to $P$, each of degree $n-1$:
${P_k(x) \stackrel{def}= \prod\limits_{i\ne k}^n (x-x_i)},\quad k = 1,\dots,n$, 
see e.g. \cite{KNUTH}, pp. 502.\FN{
They are proportional to the {\sc lagrange}-polynoms, which are defined 
as ${L_k = \prod\limits_{i\ne k}^n ((x-x_i)/(x_k-x_i))}$, i.e. $P_k(x) = \pi_k L_k(x)$, with 
$\pi_k = P_k(x_k) = \prod\limits_{i\ne k}^n (x_k-x_i)$ 
}\\
Then by definition, $P$ is also orthogonal to all $P_k$.\\
The partial polynoms $P_i$ are also mutual orthogonal ($\int w P_1P_2 = 0$). 
This is easy to see, if one expands e.g.\FN{
The $\circ$ mark stands here and sometimes in the following for omitted factors, 
to make the products more readable. Also, the subscripts $1,2$ here denote 
arbitrary, but different indices from interval $1,\dots,n$.
}
$P_1P_2 = (\circ (x-x_2)\cdots ) ((x-x_1)\circ \cdots) = P (x-x_3) \cdots$.\\
Similarly follows $\int w \; x P_1P_2 = 0$.\\

Additionally, one defines partial polynoms  $P_{kl}$ of second order (and similarly of higher order):
\begin{equation}
 \label{eq-a24}
P_{kl} \stackrel {def}= \prod_{i\ne kl} (x-x_i),\quad P_{kl} = P_{lk},
\quad P_{kk} \stackrel {def}= 0.
\end{equation}
With these definitions, I state some important equations, which can be easily proved with standard methods:
\begin{eqnarray} 
\label{eq-a25} 
P_1 P_2 = P P_{12} ,&& \qquad P_1 - P_2 = (x_1 - x_2) P_{12},\\
P' = P_1 + P_2 +\cdots, &&\qquad P_1' = P_{12} + P_{13} +\cdots = 
\sum_{i\ne 1} \frac {P_1-P_i}{x_1-x_i}. \nonumber
\end{eqnarray}

One now considers the {\em master integral} 
\begin{equation}
 \label{eq-a26}
\MAI (x_1,\cdots,x_n) \stackrel {def}= \int w P^2 = \int w (x-x_1)\cdots ^2(x-x_n)^2 > 0.
\end{equation}
 It is very easy to show, that this integral is {\em minimal} w.r.t all $x_i$:\FN{It is then
minimal among all monic polynoms.
}
At first it is stationary, since $\PD [\MAI]{x_i} = -2\int wPP_i = 0$. Secondly, it is
a real minimum, since (here I define the $n$ new constants $\rho_i \stackrel {def}= \int wP_i^2$)
\begin{equation}
 \label{eq-a27}
\PD [^2 \MAI]{x_i^2} = 2\int wP_i^2 = 2\rho_i > 0 \MBOX{and} \PD [^2 \MAI]{x_i x_j} = 0.
\end{equation}
 
One can now (uniquely) expand an arbitrary polynom $f(x)$ of degree $\le n-1$ 
by the partial polynoms $P_i$, with $n$ constants $f_i$:
\begin{equation}
 \label{eq-a28}
 f(x) = \sum_i f_i P_i(x).
\end{equation}
Then for two arbitrary polynoms $f,g$ (of degree $\le n-1$) and a linear function
$\mu = a+bx$,  easily follows (with $ \mu_i \stackrel {def}= \mu(x_i) $):\FN {
  This formula is widely used for numerical integrations, however the determination
  of the coefficients $\rho_i$ is often quite complicated. In the following,
  I will show a much more simple way to compute them, which I have not yet found in the literature.
}
\begin{equation}
 \label{eq-a29}
\int w \mu fg = \sum_i  \rho_i \mu_i f_i g_i.
\end{equation}
In fact, from this formula follows, that {\em every polynom} of degree $\le 2n-1$ can be integrated
{\em exactly} by its values at the $n$ grid points $x_i$.\\

All OP also obey a linear, second order differential equation (with $\lambda$ as eigenvalue):
\begin{equation}
 \label{eq-a30}
 u y'' + v y' +\lambda y = 0.
\end{equation}
Here $u=u_0 + u_1 x + u_2 x^2$ is a polynom of degree $\le 2$, which must not have
any zeros in the interval $[a,b]$ and $v = v_0 + v_1 x$ is linear.\FN{
By this representation, all OP systems can be easily classified.
Some important examples are: {\sc legendre}-, 
{\sc tschebyscheff}-, {\sc jacobi}-, {\sc laguerre}- and  {\sc hermite}-polynoms.
}\\ 

If then the ansatz $y = P(x) = (x-x_1)\cdots(x-x_n)$ is made and $\lambda \ne 0$,
at the zeros obviously must hold $[uy'' + vy']_{x=x_k} = 0$. With $y'(x_k) = P_k(x_k)$ and 
$y''(x_k) = 2 P_k(x_k)\sum_{i\ne k} \frac 1{x_k-x_i}$ results a system of equations for the zeros:
\begin{equation}
 \label{eq-a31}
 2\sum_{i\ne k} \frac 1{x_k-x_i} + \frac {v(x_k)}{u(x_k)} = 0.
\end{equation}
The proof, that the polynom $P= (x-x_1)\cdots(x-x_n)$ then fulfills equation (\ref{eq-a30}) is
quite simple: Since $uP'' + vP'$ is a polynom of degree $\le n$, and has (following above relations)
zeros at all $x_k$ and therefore must be proportional to $P$.\FN{
  The eigenvalue can be easily computed
  from the coefficients of $x^n$: $\lambda_n = -n(n-1)u_2  -n v_1$.
}\\
This set of equations uniquely determines the set of zeros, and also can be used to set up numerical
methods to compute them.\\
 
It also can be shown easily, that the weighting function $w(x)$ is then related to the pair 
$u,v$ by $(wu)' = wv$, i.e. $w = \frac 1u e^{\int\frac vu\; dx  }$ .\\

\subsection{Weight Factors for OP-Integrals}
As last prerequisite, a general, explicit formula for the weighting constants $\rho_i$ 
is needed (which I did not found in any textbooks). To get it, I start with the expression
(again with $u_i \stackrel {def}= u(x_i)$):\FN{
which holds for any polynom $u(x)$ of degree $\le 2$
}
\begin{equation}
 \label{eq-a32}
\Delta_{12} = \int wuP_{12}P' = \frac 1{x_2-x_1}\int wu (P_1^2 - P_2^2) = u_1\rho_1 - u_2\rho_2.
\end{equation}
On the other hand, by partial integration, one easily shows, that $\Delta_{12} = 0$, consequently
with some constant $k$, follows:\FN{The value of $k$ can be computed 
by evaluating the
integral $\int wuP'^2$, then follows $k = -(v_0 + (2n-1)v_1)\MAI $, where $v_0,v_1$ are the coefficients
of $v = v_0 + v_1x$.}
\begin{equation}
 \label{eq-a33} \underline{ u_1\rho_1 = u_2\rho_2 = \cdots = k = const.} 
\end{equation}

\subsection{Solving the Radial {\sc dirac}-Equations with OP}
With the help of above relations, the extremal principles can be investigated. Here I consider,
what one may call ``dual eigenvalue'' problems, e.g. of the type of the pair of radial {\sc dirac} equations, 
where there are two functions to variate independently\FN{
  Also {\sc lagrang}ians of 
  ordinary, second order differential equations can be represented in this form. Consider for example the 
  simple integral $ \LAGR (y) = \int (y')^2 + ay^2 \to extr.$, which leads to $y'' = ay $. 
  The last eq. is equivalent to the first order pair $y' = u, u' = ay$, 
  which in turn is represented by the {\sc lagrang}ian 
  $\LAGR (y,u) = \int yu' - uy' - ay^2 + u^2$.
} (the name of the variable  $x$ is changed from here on to $r$ and the intervall to use, is of course,
$r \in [0,\infty ]$).
The \Lagrangian here has the general form (where $a,b,c$ are some fixed functions of $r$).\FN{
  With standard variational methods, using $\int (f'g + g'f) = [fg]_0^{\infty} = 0 $, 
  one easily proves, that it is equivalent to the pair of
  first order DGL:  $f' + af + cg = 0$ and $g' - ag - bf = 0$.
}
\begin{equation}
 \label{eq-a34}
 \LAGR(f,g) = \int_0^\infty dr (f'g - g'f + 2afg + bf^2 + cg^2) \to extr.
\end{equation}
For the {\sc dirac} equation one has to use
$a = -\frac \kappa r,\;  b = \varepsilon + \frac \alpha r - m,\; c = \varepsilon + \frac \alpha r + m$, 
see e.g. \cite{LL}.

Now the polynomial ansatz is made, with a common factor $\varphi(r)$: 
${f = \varphi F}$, ${ g = \varphi G}$, where $F,G$ shall
be polynoms of degree $n-1$ (it is presumed to be possible, here), resulting in:\FN{
The derivations of $\varphi$ cancel out.}
\begin{equation}
 \label{eq-a35}
 \LAGR(F,G,\varphi) = \int_0^\infty dr \varphi^2(F'G - G'F + 2aFG + bF^2 + cG^2) \to extr.
\end{equation}
Since $ra,rb,rc$ are linear expressions of $r$, one uses the weighting function $w(r) = \varphi^2/r$,
so the factor in the integrand becomes a polynom of degree ${2(n-1) + 1 = 2n-1}$, and the above
apparatus can be applied:
\begin{equation}
 \label{eq-a36}
 \LAGR = \int_0^\infty dr w[r(F'G - G'F + 2aFG + bF^2 + cG^2)].
\end{equation}

The polynoms $F,G$ then are represented as ${F(r) = \sum_1^n f_i P_i}$, ${ G(r) = \sum_1^n g_i P_i}$,
with $2n$ constants $f_i,g_i$. 
One now uses for the derivations $F',G'$ the formula:
\begin{equation}
 \label{eq-a37}
 F' = \sum_i \hat f_i P_i,\MBOX{with}
 \hat f_i \stackrel {def}= \sum_{j\ne i} \frac {f_i+f_j}{r_i-r_j}.
\end{equation}
Inserting this all in eq. (\ref{eq-a36}), leads to a double sum:
\begin{equation}
 \label{eq-a38}
 \LAGR = \int_0^\infty dr w\sum_{ij} 
 [r(\hat f_i g_j - \hat g_i f_j + 2af_ig_j + bf_if_j + cg_ig_j)P_i P_j].
\end{equation}
If now the previously free variables $\{r_i\}$ are set to the zeros of the OP for the weighting function
$w= \varphi^2/r$, then all
integrals can be computed, and the expression becomes a simple sum 
(again defining $a_i = a(r_i),...$):
\begin{equation}
 \label{eq-a39}
 \LAGR =  \sum_{i} \rho_i r_i(\hat f_i g_i - \hat g_i f_i + 2a_if_ig_i + b_if_i^2 + c_ig_i^2).
\end{equation}

It can be shown, that the OP to use here, are {\sc laguerre}-polynoms, i.e. they belong to a 
weight function $w(r) = e^{-2\lambda r} r^{2\gamma-1}$, where 
$\lambda \stackrel {def}= \sqrt{m^2-\varepsilon^2}$ 
and $\gamma \stackrel {def}= \sqrt{\kappa^2 - \alpha^2}$.\FN{The ansatz-factor 
function $\varphi(r)$ is then $\varphi = e^{-\lambda r} r^{\gamma}$.
}
 For these, 
one has $u(r) = r$, so from eq. (\ref{eq-a33}) follows $\rho_i r_i = const. \stackrel {def}= k $ 
and I finally get the formula:
\begin{equation}
 \label{eq-a40}
\underline{ \LAGR =  k (\sum_{i \ne j} \frac{f_i g_j - g_i f_j}{r_j-r_i} + 
 \sum_i (2a_if_ig_i + b_if_i^2 + c_ig_i^2))}.
\end{equation}
This is exactly (except the constant factor $k$) the same as (\ref{eq-38}), q.e.d.\\

Again, the similarity to the starting point (\ref{eq-a34}) is remarkable, however, 
a general discussion of the preconditions for this, should be left to interested mathematicians.


\section{{\sc Lagrangian} for the Unit Sphere}
\label{sec-lag-sphere}
Here I will shortly derive the discretization scheme for the unit sphere, that leads to the
above discussed solutions of {\sc dirac} equation for the atom, in section \ref{sec-general-grid}.\FN{
  Of course, also a more concise quaternionic representation of the following is
  feasible. However, since probably most readers are not very familiar with this
  formalism, I prefer the matrix notation here.
}\\

The {\Lagrangian} for the angular part ($\MV p_i$ are the points on the unit sphere), 
that was derived there, is 
(see (\ref{eq-36})): 
\begin{equation}
  \label{eq-a41}
  \LAGR = \sum_{ij} \TR{\bar A_i \MV p_i(\MV p_i - \MV p_j)^{-1}  A_j} - 2\kappa \sum_i |A_i|,
\end{equation}
where the $A_i$ are quaternionic matrices ($\bar A_i = \HC A_i$), attached to the points
and the double sum is to build over all edges ($\MV p_i,\MV p_j$).\\

As shown below, the described grid, together with $A_i$ is {\em stationary}, i.e. 
it makes the {\Lagrangian}, considered as function $\LAGR(\MV p_i, A_i)$ extremal.\\

Any point on the unit sphere $(x,y,z)$, $x^2+y^2+z^2 = 1$ is represented with spherical 
coordinates $\vartheta,\varphi$  by the matrix
\[ \MV p =  { z, \; \; x-iy\choose x+iy, -z} = 
   {\cos \vartheta, \;\;\; \sin \vartheta e^{-i\varphi}\choose  
     \sin \vartheta e^{i\varphi}, -\cos \vartheta} .
\]
Now one introduces the  conformal one-to-one mapping of the sphere to the complex plane\FN{
  This map is the inverse of the usual
  ``{\sc riemann} sphere'' map. Please note, that the complex number $\chi$ then transforms 
  {\em  linear fractionally} under ordinary space-rotations, namely if 
  $T = {\alpha, \;\; \beta \choose -\CC \beta, \CC \alpha}$ is a rotation (with $|T|=1$), 
  then follows $\chi \to \frac {\alpha \chi +\beta}{-\CC\beta \chi + \CC\alpha}$.
} $\chi \in {\bf C}$ with:\FN{
  E.g. the north pole $(0,0,1)$ is mapped to the origin of the complex plane $\chi = 0$, 
  the south pole  $(0,0,-1)$  to the infinite point $\chi = \infty$ and the equator to
  the circle $|\chi| = 1$.
}
\begin{equation}
  \label{eq-a42}
  \chi = \tan \frac \vartheta 2 e^{i\varphi}= \frac {\sin\vartheta  e^{i\varphi}}{1 + \cos \vartheta}.
\end{equation}
Then one gets with simple trigonometric identities: 
$z = \cos \vartheta = \frac {1-|\chi|^2}{1+|\chi|^2} $ and 
$x+iy = \sin \vartheta e^{i\varphi} = \frac {2 \chi}{1+|\chi|^2} $, consequently:
\[\MV p = \frac 1{1+|\chi|^2} {1-|\chi|^2, 2 \CC\chi\choose 2\chi, |\chi|^2 -1 }.
\]
This matrix can be decomposed with the help of {\em spinor-matrix factors}\FN{
  These factors transform similar to spinors under rotations, not {\sc minkowski} space vectors!
} $Q = Q(\chi)$, which I define here as:
\begin{equation}
  \label{eq-a43}
   Q(\chi) \DEF { 1,\;\; \CC\chi \choose \chi, -1}
\end{equation}
and the constant matrix $U \DEF {1,\;\; 0\choose 0,-1}$, namely as:\FN{
  For an explicit proof consider $| Q| = -(1+|\chi|^2)$, $ Q^2 = 1+|\chi|^2$
  and $ Q ^{-1} = \frac {1}{1+|\chi|^2}  Q$
}
\begin{equation}
  \label{eq-a44}
  \MV p = Q U  Q^{-1}.
\end{equation} 
With the help of this decomposition formula it is easy to express the required 
difference vector of {\em two arbitrary points} $\MV p_1, \MV p_2$ on the unit sphere (the
inverse of this difference is the crucial part in computing the 
{\Lagrangian} sum in (\ref{eq-a41})) and $Q_i \DEF Q(\chi_i)$:\FN{
  Consider from above $ Q_k^{-1} U  Q_k =  Q_k U  Q_k^{-1}$
}
\begin{equation}
  \label{eq-a45}
  \MV p_1- \MV p_2 =  Q_1U  Q_1^{-1} -   Q_2U  Q_2^{-1} =
   Q_2^{-1} \underbrace{\left(  Q_2 Q_1 U -   U Q_2 Q_1\right)}_{\DEF D_{21}} Q_1^{-1} 
\end{equation}  
With  $V_{21} \DEF  Q_2 Q_1 = 
{1 + \CC\chi_2\chi_1, \CC\chi_1 - \CC\chi_2\choose \chi_2-\chi_1, 1 + \chi_2\CC\chi_1}$ 
it is to see, that the  difference 
in the brackets of this expression  $D_{21}$, 
is the anticommuting part of the factors $U$ and $V_{21}$ and can easily be 
computed as:
\[ D_{21} = 2{0, \CC\chi_2 - \CC\chi_1 \choose \chi_2-\chi_1, 0}.
\]
With this equation (\ref{eq-a45}) is easily invertable and one finally gets:
\begin{equation}
  \label{eq-a46}
  \MV p_1 (\MV p_1 - \MV p_2)^{-1} = Q_1 U Q_1^{-1} Q_1 D_{21}^{-1} Q_2 = 
  \frac 12  Q_1 
  {0, \frac 1{\chi_2 - \chi_1}\choose - \frac 1{\CC\chi_2-\CC\chi_1}, 0} Q_2.
\end{equation}
Now one makes the following substitution for the $A_i$, to simplify the double sum in 
(\ref{eq-a41}), with the complex constants $\mu_i,\nu_i$:\FN{
  This is the most general ansatz, if the quaternionic restriction $\bar A_i = \HC A_i$ 
  is considered.
}
\begin{equation}
  \label{eq-a47}
  A_i = Q_i^{-1} {\CC \mu_i, \;\;\ \nu_i \choose \CC\nu_i, - \mu_i}. 
\end{equation}
Then follows 
$|A_i| = \frac 1{|Q_i|} (-\mu_i \CC \mu_i -\nu_i \CC \nu_i ) = 
\frac {|\mu_i|^2 + |\nu_i|^2}{1+|\chi_i|^2}$.\\
The {\Lagrangian} in (\ref{eq-a41}) then  simplifies to (it is obviously real as required):
\begin{equation}
  \label{eq-a48}
  \LAGR(\mu_i,\nu_i,\chi_i) = \sum_{ij}\left(\frac {\CC \nu_j \mu_i}{\chi_j -\chi_i} +
    \frac { \nu_j \CC \mu_i}{\CC \chi_j -\CC\chi_i}\right) -
  2\kappa \sum_i\frac{|\mu_i|^2 + |\nu_i|^2}{1+|\chi_i|^2}.
\end{equation}
As always, one has to find {\em stationary points} in parameter space, which is here 
$\{\mu_i,\nu_i,\chi_i\}, i = 1,\dots, n$.\\

Since all parameters are simple scalars, this can be done by setting the partial derivations zero,
which gives $3n$ equations, $k=1,\dots n$:\FN{
  Us usual the complex conjugate of any parameter can be considered
  as independent, and since the expression $\LAGR$ is real, 
  the derivation by it leads to an equivalent equation. 
} 
\[ \PD[\LAGR]{\mu_k} = \PD[\LAGR]{\nu_k} = \PD[\LAGR]{\chi_k} = 0
\]  
Namely, (by the variations of $\CC \nu_i, \CC \mu_i$) result the first $2n$ equations 
(for $k = 1,\dots,n$):\FN{
  Consider the complex differentiation rules, e.g.
  $\PD[|\chi|^2] \chi = \PD[(\chi \CC \chi)] \chi = \CC \chi$.
}
\begin{equation}
  \label{eq-a49}
  \sum_i \frac {\mu_i}{\chi_k-\chi_i} \stackrel != 2\kappa \frac {\nu_k}{1+|\chi_k|^2}\MBOX{and}
  \sum_i \frac {\nu_i}{\CC\chi_k-\CC\chi_i} \stackrel != -2\kappa \frac {\mu_k}{1+|\chi_k|^2}.
\end{equation}
The variation of all $\chi_k$ leads to the  equations ($k = 1,\dots,n$):\FN{The double sum
  contains each pair $i,k$ {\em twice}.
}
\begin{equation}
  \label{eq-a50}
  -\sum_i \frac {\CC \nu_k\mu_i- \CC \nu_i \mu_k}{(\chi_k-\chi_i)^2} + 
  2\kappa \frac {(|\mu_k|^2 + |\nu_k|^2) \CC \chi_k}{(1+|\chi_k|^2)^2} = 0.
\end{equation}

Now I will show, that the point in parameter space, given by the simple formulas
\underline{$\mu_i = c $, $\nu_i = c \CC \chi_i$} with real constant $c$\FN{
  The resulting matrix is then simply
  $A_i = Q_i^{-1} c Q_i = cI$. 
}
and any set of points $\{\chi_k\}$ (on the unit sphere by definition), 
that obeys
\begin{equation}
  \label{eq-a51}
  \sum_i \frac {1+\chi_i\CC\chi_k}{\chi_k -\chi_i} \stackrel != 0,\quad k = 1,\dots, n
\end{equation}
fulfills all $3n$ conditions above.
Like all others, the above sum is {\em not to build over all} $i \ne k$,
but only over the edges $(\chi_i,\chi_k)$, which are presented in the following.\FN{
  These equations are indeed {\em covariant} under space rotations, described here as linear fractional
  transformation of all $\chi_i$, like mentioned above.

  The simplest possible example set is given with $n=2$, i.e. two points $\chi_1,\chi_2$ 
  with one edge between them.
  Both (\ref{eq-a51}), for $k=1,2$, then simply require $1+\chi_1\CC\chi_2 \stackrel != 0$. 
  This condition says, 
  that the two points must be {\em antipodes} on the sphere, 
  while one of them, e.g. $\chi_1$, is freely variable.
}
This grid of points $\chi_k$ can, as to expect, be derived from {\em spherical harmonics} 
(this is described below).

Simple rearrangement gives two equivalent sets of equations, wherein $l$ is the number of
summands (edges $i,k$), which should be equal for all $\chi_k$, $k = 1,\dots, n$:
\begin{equation}
  \label{eq-a52}
  \sum_i \frac {1}{\chi_k -\chi_i} = l \frac {\CC \chi_k}{1+|\chi_k|^2} \MBOX{and}
  \sum_i \frac {\chi_i}{\chi_k -\chi_i} = -l \frac {1}{1+|\chi_k|^2}.
\end{equation}

Then all $3n$ conditions (\ref{eq-a49}, \ref{eq-a50}) are fulfilled,  
i.e. {\bf the stationarity of $\LAGR(\mu_i,\nu_i,\chi_i)$ is proven}, 
if the eigenvalue $\kappa$ is set to 
\underline{$\kappa = \frac l2$}, q.e.d.\FN{
  which is to see by simply inserting the above ansatz: 
  $\mu_i = 1, \nu_i = \CC\chi_i$ and (\ref{eq-a52}).
}\\

A second stationary point is obviously given with the same set of $\chi_k$, but\FN{
  The resulting matrix is then 
  $A_k = Q_k^{-1} c {\CC \chi_k, 1\choose 1,-\chi_k} = 
  \frac c{1+|\chi_k|^2} {2\CC\chi_k, 1-|\chi_k|^2\choose |\chi_k|^2-1, 2\chi_k } =
        c { x_k -iy_k, z_k \choose -z_k, x+iy_k} $
}
\[ \mu_i = c\chi_i, \quad \nu_i = c, \quad \kappa = -\frac l2.
\]
Please note, that both cases $\kappa = \pm \frac l2$ describe {\em different quantum states} in the
{\sc dirac} equation (see e.g. \cite{LL}, pp. 119).\\
Finally I will shortly sketch, how a set of $\{\chi_k\}$ and the assigned edges, that 
fulfill (\ref{eq-a51}), 
can be constructed using {\em spherical harmonics}.\\
There might exist also other grids, which give the same result, but I was not able to find anyone.\\

The nodes are supposed to be arranged on $h$ latitude circles 
(which are defined here by $|\chi| = const.$), in the way that on every circle
are $2m$ equidistant nodes.\\ 
I.e. one has $n = 2m\times h$ nodes and they can be assigned as (with real $t_k > 0$):
\begin{equation}
  \chi = t_k e^{\pi i \frac jm},\quad k=1,\dots h,\quad j = 1,\dots 2m.
\end{equation}
Edges shall be only on latitude circles (denoted as $i \in B(k)$) and longitude circles 
(as $i\in A(k)$):
\begin{equation}
  \sum_i \frac 1{\chi_k -\chi_i} =  \sum_{i \in B(k)} \frac 1{\chi_k -\chi_i} + 
  \sum_{i \in A(k)}  \frac 1{\chi_k -\chi_i}.
\end{equation}
This gives a total number of $l = 2m-1 + 2h-1 = 2(m+h-1)$ edges connected to every node, 
since on each longitude circle $j$ are 
actually $2h$ nodes: $t_k e^{\pi i\frac jm }$ and $t_k e^{\pi i\frac {j+m}m } = -t_k e^{\pi i\frac jm}$, 
$k = 1,\dots h$. This also means, that the opposite point of $\chi_k$ on the latitude circle,
which is $-\chi_k$, counts {\em twice} in the sum (on both circles).\\

The summation on a latitude circle gives:\FN{
  This formula can be proved using a general relation for the complex roots of $z^n = 1$, which are
  ${z_k = e^{2\pi ik/n}}, \;\; k=1,\dots,n$, namely:
  $\sum_{k=1}^{n-1} \frac 1{1-z_i} = \frac {n-1}2$, which again follows from 
  $\frac 1{1-z} + \frac 1{1-1/z} = 1 $.
}
\begin{equation}
  \sum_{i \in B(k)} \frac 1{\chi_k -\chi_i} = \frac {2m-1}{2\chi_k}.
\end{equation}
The summation on the longitude circle gives (the opposite point of $\chi_k$ on this circle, 
gives the summand $1/2\chi_k$): 
\begin{equation}
  \sum_{i \in A(k)} \frac 1{\chi_k -\chi_i} =    \frac 1{2\chi_k} +
  \frac {t_k}{\chi_k} \Bigg[\sum_{i \ne k}  (\frac 1{t_k - t_i} + \frac 1{t_k + t_i})\Bigg].
\end{equation}
Consequently the complete sum over all edges becomes:
\begin{equation}
  \label{eq-a57}
  \sum_i \frac 1{\chi_k -\chi_i} = 
  \frac {\CC\chi_k}{t_k^2} \Big[ m + \sum_{i \ne k} \frac {2t_k^2}{t_k^2 - t_i^2} \Big].
\end{equation}
To prove corespondence to zeros of spherical harmonics, one subtitutes back to the 
cartesian coordinates, which is given by $t^2 = |\chi|^2 = \frac {1-z}{1+z}$. Then (\ref{eq-a57}) 
can be expressed as 
\[
\sum_i \frac 1{\chi_k -\chi_i} = 
\CC\chi_k (1+z_k) \Big[\frac m{1-z_k} + \sum_i \frac {1+z_i} {z_i-z_k}\Big].
\]
The r.h.s. of this equation is then equal to the required expression 
\[
l\frac {\CC \chi_k}{1+|\chi_k|^2} = l \CC\chi_k \frac {1+z_k}2 = (m+h-1) \CC\chi_k(1+z_k)
\]
to obey the stationarity conditions (\ref{eq-a52}), if for all $z_k$  holds:
\begin{equation}
  \label{eq-a58}
\sum_i \frac 1{z_k-z_i} = m \frac {z_k}{1-z_k^2}.
\end{equation}

It is now easy to check, that (\ref{eq-a58}) is fulfilled
for the \underline {zeros of {\sc legendre} functions} of order $m+h$, namely
$P^{m-1}_{m+h}(z)$ (see e.g. \cite{COUR-HILB}, pp. 282), with the methods of orthogonal polynoms
presented here (see section \ref{sec-orth-poly}, eq. (\ref{eq-a31})).\\

As a summary of this chapter, I want to state, that all classical stationary
states of the {\sc dirac} equation are reproduced exactly above. 
However, it is not yet clear, by which
principle some grids (e.g. with
an odd number of nodes on a latitude circle) are suppressed.

\end{appendix}


\begin{thebibliography}{99}

\bibitem{BELL} J.S. Bell,  {\sl Speakable and unspeakable in quantum mechanics},
 University Press,  Cambridge 1987.

\bibitem{BJORKEN} J.D. Bjorken, S.D. Drell, {\sl Relativistische Quantenmechanik}, 
Bibliographisches Institut, Mannheim 1966.

\bibitem{COUR-HILB} R. Courant, D. Hilbert, {\sl Methoden der mathematischen Physik}, 
Vierte Auf\-lage, Springer Verlag, Berlin Heidelberg, 1993 

\bibitem{EBERT}  Dietmar Ebert, {\sl Eichtheorien}, Akademie Verlag, Berlin 1989. 

\bibitem{EINSTEIN}  Albert Einstein, {\sl Grundzüge der Relativitätstheorie}, Akademie Verlag, 
Berlin 1973. 

\bibitem{HS-SRT}  E. Herlt, N. Salie, {\sl Spezielle Relativitätstheorie}, Akademie Verlag, Berlin 1978. 

\bibitem{KNUTH} Donald E. Knuth,  {\sl The art of computer programming}, 
Vol. 2 : ``Seminumerical Algorithms'',
Addison-Wesley, Boston  1998.

\bibitem{LL} Landau, Liftschitz, {\sl Quanten-Elektrodynamik}, Akademie Verlag,
Berlin 1991.

\bibitem{PENROSE} Roger Penrose, Wolfgang Rindler, {\sl Spinors and space-time}, Vol. 1, Cambridge
University Press, Cambridge 1984.

\bibitem{PENROSE-EMP} Roger Penrose, {\sl The Emperor's New Mind}, Oxford
University Press, 1989.

\end{thebibliography}
\end{document}